\documentclass[superscriptaddress,secnumarabic,
amssymb,amsmath,nobibnotes,aps,prd,showpacs,nofootinbib]{revtex4}%
\usepackage{graphicx}
\usepackage{epsf}
\usepackage{bm}
\usepackage{amsmath}
\usepackage{amsfonts}
\usepackage{amssymb}
\usepackage{epstopdf}
\usepackage{natbib}
\usepackage{color}%
\usepackage{float}
\providecommand{\U}[1]{\protect\rule{.1in}{.1in}}
\newcommand{\be}{\begin{equation}}
\newcommand{\ee}{\end{equation}}

\newcommand{\mincir}{\raise
-3.truept\hbox{\rlap{\hbox{$\sim$}}\raise4.truept\hbox{$<$}\ }}
\newcommand{\magcir}{\raise
-3.truept\hbox{\rlap{\hbox{$\sim$}}\raise4.truept\hbox{$>$}\ }}

\newtheorem{remark}{Remark}[section]

\begin{document}



\title{Quintessential inflation at low reheating temperatures}


\author{Llibert Arest\'e Sal\'o\footnote{E-mail: llibert.areste@estudiant.upc.edu, Llibert.Salo@campus.lmu.de} }
\affiliation{Departament de Matem\`atiques, Universitat Polit\`ecnica de Catalunya, Diagonal 647, 08028 Barcelona, Spain}
\affiliation{Fakult{\"a}t f{\"u}r Physik, Ludwig-Maximilians-Universit{\"a}t, Theresienstr. 37, 80333 M{\"u}nchen, Germany}

\author{Jaume de Haro\footnote{E-mail: jaime.haro@upc.edu}}
\affiliation{Departament de Matem\`atiques, Universitat Polit\`ecnica de Catalunya, Colom 11, 08222 Terrassa, Spain}


\thispagestyle{empty}

\begin{abstract}
We have tested some simple quintessential inflation models, imposing that they match with the recent observational data provided by the BICEP and Planck's team and leading to a reheating temperature, which is obtained via gravitational particle production after inflation, supporting the nucleosynthesis success. { Moreover, for the models coming from supergravity one needs to
demand low temperatures in order to avoid problems such as the gravitino overproduction or the gravitational production of moduli fields, which are obtained only when the reheating temperature is due to the production of massless particles with a coupling constant very close to its conformal value.}

\end{abstract}

\vspace{0.5cm}



\maketitle

\section{Introduction}

Quintessential inflation  \cite{dv} is a good candidate to unify the early and late time acceleration of our universe. These models are a combination of an inflationary potential, used to explain the early acceleration of the universe, and a quintessential one -that could be only a cosmological constant-, which takes into account the current cosmic acceleration. At early times the inflationary acceleration is the one that dominates and it ceases to be dominant in a phase transition where the universe enters in a kination regime \cite{Joyce}. At this moment, particles coupled with gravity are produced to reheat the universe and match with the current hot one. Finally, at very late times, the quintessential potential dominates and the universe starts to accelerate again.

\

In order to assure the viability of these models, they are required to fit well with the recent observational data provided by the BICEP and Planck's teams \cite{bicep2, Planck}, but also the reheating temperatures have to be compatible with the nucleosynthesis and baryogenesis bounds, preventing the increase of entropy due to the decays of gravitational relics such as moduli fields and gravitinos.
It is well known that the success of nucleosynthesis constraints the reheating temperature to be between $1$ MeV and $10^9$ GeV \cite{Allahverdi, kl, gkr,  eln, nos, ekn, klss}, whereas in order to overcome the overproduction of very light gravitinos { (particles which appear in supersymmetric gravitational theories)}, i.e., with masses in the range $ 10^{-3}$ MeV $\lesssim  m_{3/2}\lesssim 10^{-1}$ MeV, it is stated that the reheating temperature has to be less than $10^2$ GeV \cite{kty}. It is more complicated to avoid the increase of entropy due to the production of moduli fields, { which also appears in supergravity models}, during the inflationary period; in that case, as it was showed in \cite{fkl}, the reheating temperature has to be less than $1$ GeV. However, this low temperature seems to be problematic with baryogenesis, according to which it is commonly assumed that the reheating temperature is at least of the order of the electroweak scale ($10^2$ GeV). But, fortunately, there are mechanisms that explain the baryon asymmetry 
at very low temperatures \cite{dlr}. For all these reasons, 
we will consider that those { supergravity} models with a reheating temperature in the MeV regime, i.e., between $1$ MeV and $1$ GeV, succeed in solving all these problems. 


\ 

Having this in mind, 
the main goal of the present work is to study the viability of some well-known inflationary potentials, {  some of them coming from supersymmetric theories},  adapted to quintessence. To do it in a simple way, first of all we consider a universe with a cosmological constant, we choose positive inflationary potentials that vanish at some value of the scalar field and then we extend them to zero for the other values of the field. Hence, we obtain a potential with a phase transition that models an inflationary universe at early times, which reheats the universe via gravitational particle production after inflation and it is finally dominated by the cosmological constant. Once we have these potentials, we calculate their spectral parameters, the number of efolds, which has to be in quintessential inflation between 63 and 73 as we show, and its reheating temperature in three cases that will be analytically calculated: via the production of heavy massive particles conformally coupled with gravity, massless nearly conformally coupled particles and particles far from the conformal coupling. This last case has also been studied in \cite{sss} in the context of braneworld inflation for exponential and power law potentials. Note that there is another way to reheat the universe, via the so-called ``instant preheating" \cite{fkl}, which has been applied to quintessential inflation in  \cite{ghmss, hmss} for exponential and power law potentials. This kind of reheating deserves future investigation for different potentials such as the ones studied in this paper, which we will deal with in a future work.

\

The paper is organized as follows: In Section II we build a non singular model based on a universe filled with a barotropic fluid with a non-linear equation of state where the universe starts and finishes in a de Sitter phase. Via the reconstruction method we find the potencial that mimics these dynamics when the universe is filled with a scalar field, we find the analytic solution that mimics the 
fluid model, and we see that it provides spectral quantities such as the spectral index, its running and the tensor to scalar ratio that match with the current observational data. However, this solution appears to be unstable at early times, in the sense that any small perturbation leads to a solution which is singular at early times, in the same way that happens with the Starobinsky model \cite{starobinsky, aho}. Moreover, with our criterion, the model only supports a reheating via the creation of massless particles nearly conformally coupled with gravity.
In Section III, we consider a universe with a small cosmological constant and we adapt three simple inflationary potentials (Exponential SUSY Inflation potential (ESI), the Higgs Inflation Potential in Einstein Frame (HI), and the Power Law Inflation potential  (PLI))  to quintessence, showing that the HI potential leads to an unacceptable number of efolds, the ESI potential, { which comes from a supersymmetric potential and thus suffers the gravitino and moduli problems, only supports a reheating via the production of massless particles nearly conformally coupled with gravity}, and the PLI, { which is only viable in the quadratic case,
supports the three kind of reheating studied in the work because, since it does not come from any supergravity potential, gravitinos or relic moduli fields do not appear in this theory}. In the last Section we adapt other well known potentials 
 { (Witten-O'Raifeartaigh Inflation (WRI), K\"ahler Moduli Inflation I (KMII), 
 Open String Tachionic Inflation (OSTI), 
 Brane Inflation (BI) and Loop Inflation (LI))} coming from inflation,  to quintessence and we study its viability. { Finally, in the appendix we review from a critical viewpoint the overproduction of gravitational waves in quintessential inflation}.

\vskip 0.3cm

The units used throughout the paper are $\hbar=c=1$ and, with these units, $M_{pl}=\frac{1}{\sqrt{8\pi G}}$ is the reduced Planck's mass.

\section{A nonsingular model}

In this Section we are going to study  inflation coming from fluids \cite{noo,bnos} and
 to develop the idea of a nonsingular universe proposed in \cite{ha}, where the universe was filled with a barotropic fluid whose non-linear Equation of
State (EoS), namely $P=P(\rho)$ being $P$ the pressure and $\rho$ the energy density, satisfies $P+\rho=0$ at two different scales, leading, at early and late times, to two de Sitter eras.

\

The simplest realization of this model, which could be generalized introducing viscosity \cite{bghos},  is to consider
three parameters $H_i\gg H_E \gg H_f$,
where $H_i$ and $H_f$ are two fixed
points of the dynamical system and $H_E$ is the value of the Hubble parameter when the transition from inflation to kination is produced. To simplify, we assume that
$H_i=M_{pl}$, i.e., the universe starts, in a de Sitter phase, at Planck scales, and in order to reproduce the current cosmic acceleration we have to take $H_f\sim H_0$. The simplest way to obtain the model is to consider the following differential equation

\begin{eqnarray}\label{dynamics}
\dot{H}=\left\{\begin{array}{ccc}
-k(M_{pl}-H)^2& \mbox{for}& H\geq H_E\\
-3(H-H_f)^2 &\mbox{for}& H\leq H_E,
\end{array}\right.
\end{eqnarray}
where we have chosen $k=\frac{3(H_E-H_f)^2}{(M_{pl}-H_E)^2}\cong \frac{3H_E^2}{M_{pl}^2}$ to ensure the continuity of $\dot{H}$. Regarding the value of $k$, it will be determined in next section so that our model matches with the current observational data. The dynamical system can be analytically solved leading to the following Hubble parameter 
\begin{eqnarray}
 H(t)=\left\{\begin{array}{ccc}
      \frac{H_E-ktM_{pl}(M_{pl}-H_E)}{1-kt(M_{pl}-H_E)}&\mbox{when}& t\leq 0\\
      \frac{H_E+3H_ft(H_E-H_f)}{1+3t(H_E-H_f)} &\mbox{when}& t\geq 0,
             \end{array}\right.
\end{eqnarray}
and the corresponding scale factor is
\begin{eqnarray}
 a(t)=\left\{\begin{array}{ccc}
      a_E (1-kt(M_{pl}-H_E))^{1/k}e^{M_{pl}t}
       & \mbox{when}&t\leq 0\\
    a_E (1+3t(H_E-H_f))^{1/3}e^{H_ft}  & \mbox{when}&t\geq 0.
             \end{array}\right.
\end{eqnarray}

Moreover,
the effective EoS parameter, namely $w_{eff}$, which is defined as $w_{eff}\equiv \frac{P}{\rho}=-1-\frac{2\dot{H}}{3H^2}$, for our  model is given by
\begin{eqnarray}
 w_{eff}=\left\{\begin{array}{ccc}
      -1+\frac{2k}{3}\left(\frac{M_{pl}}{H}-1\right)^2  &\mbox{when}& H\geq H_E\\
      -1+2 \left(1-\frac{H_f}{H} \right)^2 & \mbox{when}& H\leq H_E,
             \end{array}\right.
\end{eqnarray}
which shows that for $H\sim M_{pl}$ one has
$w_{eff}(H)\cong -1$, meaning that we have an early inflationary quasi de Sitter period. At the phase transition, i.e. when $H\cong H_E$, the EoS parameter satisfies $w_{eff}(H)\cong 1$ and the universe enters in a kination or deflationary period \cite{spokoiny,joyce}, and finally, for $H\cong H_f$ one also has  $w_{eff}(H)\cong -1$ depicting the current cosmic acceleration.

\subsection{Cosmological perturbations}

In order to study the cosmological perturbations, one needs to introduce the slow roll parameters \cite{btw}
\begin{eqnarray}
\epsilon=-\frac{\dot{H}}{H^2}, \ \ \ \eta=2\epsilon-\frac{\dot{\epsilon}}{2H\epsilon},
\end{eqnarray}
which allow us to calculate the associated inflationary parameters, such as the spectral index ($n_s$), its running ($\alpha_s$) and the ratio of tensor to scalar perturbations (r) defined below

\begin{eqnarray}
n_s-1=-6\epsilon_*+2\eta_*, \ \ \ \ \alpha_s=\frac{H_*\dot{n}_s}{H_*^2+\dot{H}_*}, \ \ \ \ r=16\epsilon_*,
\end{eqnarray}
where the star ($*$) means that the quantities are evaluated when the pivot scale crosses the Hubble radius. In our case, these quantities become

\begin{eqnarray}
\frac{n_s-1}{2k}=\frac{1}{H_*^2}(M_{pl}-H_*)(2M_{pl}-H_*),   \ \ \ \ \alpha_s=\frac{-2k^2M_{pl}(M_{pl}-H_*)^2(4M_{pl}-3H_*)}{H_*^2(H^2-k(M_{pl}-H)^2)}, \ \ \ \ r=16k\left(\frac{M_{pl}}{H_*}-1\right)^2. \label{param}
\end{eqnarray}

\

From the theoretical \cite{btw} and the observational \cite{bld} value of the power spectrum
\begin{eqnarray}
 {\mathcal P}\cong \frac{H_*^2}{8\pi^2\epsilon_* M_{pl}^2}\sim 2\times 10^{-9},  \label{pot}
\end{eqnarray}
and the observational value of the spectral index obtained by Planck2015 data, $n_s=0.968\pm 0.006$, we can find the value of $H_*$ and $k$ that fits in our model with these data by inserting the value 
of $k=\frac{10^9H_*^4}{16\pi^2(M_{pl}-H_*)^2M_{pl}^2} \left( \text{coming from Eq. \eqref{pot}, using that } \epsilon_*=k\left(1-\frac{M_{pl}}{H_*}\right)^2\right)$ into the first equation in \eqref{param}.  This is equivalent to finding a root to $f(H)=\alpha(M_{pl}-H) M_{pl}^2-H^2(2M_{pl}-H)$, where $\alpha=8\times 10^{-9}\pi^2(1-n_s)>0$. Since $f(0)>0$, $\lim\limits_{H\to-\infty}f(H)<0$, $\lim\limits_{H\to\infty}f(H)>0$ and $f(M_{pl})<0$, 
from Bolzano's theorem, there is always a unique root $H_*$ such that $0<H_*<M_{pl}$.

\

Hence, for the corresponding range of values of $n_s$, we obtain that $3\times 10^{-5}<\frac{H_*}{M_{pl}}<4\times 10^{-5}$. Thus, it is straightforward that $7\times 10^{-12}<k<10^{-11}$ and, by using the equations in \eqref{param}, one gets $0.1040<r< 0.1520$ and $-7\times 10^{-4}<\alpha_s < -3\times 10^{-4}$. Therefore, some values of $n_s$ in the 1-dimensional marginalized $2\sigma$ C.L. ($0.970<n_s<0.974$) correspond to a ratio of tensor to scalar perturbations which agrees with the constraint $r<0.12$, provided by BICEP and Planck collaboration \cite{bicep2}. It fulfills, as well, the limits obtained for $\alpha_s$ in Planck2015 \cite{Planck} ($\alpha_s=-0.003\pm 0.007$). 

\

Finally, the number of e-folds is given by
\begin{eqnarray}
N=\int_{t_*}^{t_{end}} H dt =-\int_{H_{end}}^{H_*}\frac{H}{\dot{H}}dH=\frac{1}{k}\left[\ln\left(\frac{M_{pl}-H_*}{M_{pl}-H_{end}} \right)+M_{pl}\left(\frac{1}{M_{pl}-H_*}-\frac{1}{M_{pl}-H_{end}} \right) \right] \label{efolds}
\end{eqnarray}
where (end) stands for the end of inflation, i.e., $\epsilon_{end}=1$, that is, $H_{end}=\frac{M_{pl}\sqrt{k}}{1+\sqrt{k}}$. Thus, by using the calculated values of $k$ and $H_*$, we obtain a number of e-folds satisfying $52<N< 77$. Now, we are going to compare this value with the one that we will obtain from the following equation \cite{liddle} in an analogous way as in \cite{inflation2}:

\begin{eqnarray}
\frac{k_*}{a_0H_0}=e^{-N}\frac{H_*}{H_0}\frac{a_{end}}{a_E}\frac{a_E}{a_R}\frac{a_R}{a_M}\frac{a_M}{a_0}=e^{-N}\frac{H_*}{H_0}\frac{a_{end}}{a_E}\frac{\rho_R^{-1/12}\rho_M^{1/4}}{\rho_E^{1/6}}\frac{a_M}{a_0},
\end{eqnarray}
where $R$ and $M$ symbolize the beginning of radiation era and the beginning of the matter domination era and we have used relations $(a_E/a_R)^6=\rho_R/\rho_E$ and $(a_R/a_M)^4=\rho_M/\rho_R$. We use that $H_0\approx 2\times 10^{-4}\text{Mpc}^{-1}$ and, as usual, we choose $a_0=1$ taking as a physical value of the pivot scale $k_{phys}=0.02\text{ Mpc}^{-1}$
(value used by Planck2015 \cite{Planck}). Then one has that, in co-moving coordinates, the pivot scale will be $k_*\equiv a_0k_{phys}= 0.02\text{ Mpc}^{-1}$.
Moreover, we know that the process after reheating is adiabatic, i.e. $T_0=\frac{a_M}{a_0}T_M$, as well as the relations $\rho_M\approx\frac{\pi^2}{15}g_MT_M^4$ and $\rho_R\approx\frac{\pi^2}{30}g_RT_R^4$ (where $\{g_i\}_{i=R,M}$ are the relativistic degrees of freedom \cite{rehagen}). Hence,
\begin{eqnarray}
N=-4.61+\ln\left(\frac{H_*}{H_0} \right)+\ln\left(\frac{a_{end}}{a_E} \right)+\frac{1}{4}\ln\left(\frac{2g_M}{g_R}\right)+\frac{1}{6}\ln\left(\frac{\rho_R}{\rho_E} \right)+\ln\left(\frac{T_0}{T_R} \right).
\end{eqnarray}

We use that $H_0\sim 6\times 10^{-61}M_{pl}$ and, from Equation \eqref{pot}, we infere that $H_*\sim 4\times 10^{4}\sqrt{\epsilon_*}M_{pl}$. We know as well that $T_0\sim 2\times 10^{-13} \text{GeV}$ and $g_M=3.36$ \cite{rehagen}. Also, $g_R=107$, $90$ and $11$ for $T_R\geq 135$ GeV, $175 \text{ GeV}\geq T_R\geq 200$ MeV and $200\text{ MeV}\geq T_R\geq 1$ MeV, respectively \cite{rehagen}. On the other hand,
\begin{eqnarray}\label{eq 12}
\ln\left(\frac{a_{end}}{a_E} \right)=\int_{H_E}^{H_{end}}\frac{H}{\dot{H}}dH\approx-\frac{1}{k}\left[\frac{1}{1-\sqrt{k}}-\frac{1}{1-\sqrt{\frac{k}{3}}}+\ln\left(\frac{1-\sqrt{k}}{1-\sqrt{\frac{k}{3}}} \right) \right]
\cong -0.3.
\end{eqnarray}

Thus, we obtain that
\begin{eqnarray}
N\approx 54.5+\frac{1}{2}\ln\epsilon_*
-\frac{1}{3}\ln\left(\frac{g_R^{1/4}T_RH_E}{M_{pl}^2}\right). \label{eq13}
\end{eqnarray}

Therefore, with the values in our model and with the range $1\text{ MeV} \leq T_R\leq 10^9$ GeV required in order to have a successful nucleosynthesis \cite{giudice}, we find that $63\lesssim N \lesssim 73$. With equation \eqref{efolds}, this range is verified in our values for $0.969<n_s<0.973$.

\

\begin{figure}[H]
\begin{center}
\includegraphics[scale=0.48]{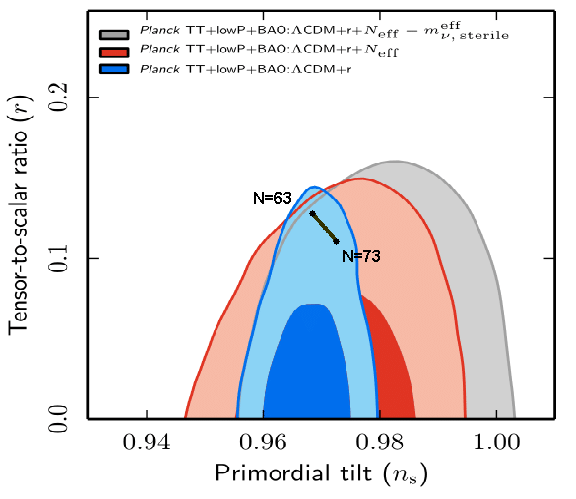}
\end{center}
\caption{Marginalized joint confidence contours for $(n_s, r)$, at the 68\% and 95\% CL, without the presence of running of the spectral indices. We have drawn the curve from $N=63$ to $73$ e-folds. (Figure courtesy of the Planck2015 Collaboration).}
\label{fig:planck}
\end{figure}

In conclusion, with the intersection of the bounds obtained from the constraints $r<0.12$ and $63\lesssim N \lesssim 73$, in our model we have that $0.970<n_s<0.973$ with the corresponding values of $k\sim 8\times 10^{-12}$ and $H_*\sim 3\times 10^{-5}M_{pl}$ (corresponding to $H_E\sim 2\times 10^{-6}M_{pl}$), thereby having a number of e-folds of $66\leq N\leq 73$. Hence, our model satisfies all the values obtained from recent observations. Moreover, one can see in Figure \ref{fig:planck} that our model also provides theoretical values that enter in the marginalized $2\sigma$ C.L. contour in the plane $(n_s,r)$.

\subsection{The scalar field}

In this section we are going to mimic the perfect fluid that fills the FLRW universe with a scalar field. If we represent the energy density and the pressure by the notations $\rho_{\varphi}$, $p_{\varphi}$, respectively, then they assume the following simplest forms:
\begin{eqnarray}
\rho_{\varphi}=\frac{\dot{\varphi}^2}{2}+V(\varphi), \ \ \ \ p_{\varphi}=\frac{\dot{\varphi}^2}{2}-V(\varphi). \label{prho}
\end{eqnarray}

Now, using Eq. \eqref{prho} and the Raychaudhuri equation $\dot{H}=-\frac{\dot{\varphi}^2}{2M_{pl}^2}$, we find
\begin{eqnarray}
\varphi=-M_{pl}\int\sqrt{-2\dot{H}}dt=-M_{pl}\int\sqrt{-\frac{2}{\dot{H}}}dH,
\end{eqnarray}
which can be analytically solved

\begin{eqnarray}
\varphi=\left\{\begin{array}{ccc}
M_{pl}\sqrt{\frac{2}{k}}\ln\left(\frac{M_{pl}-H}{M_{pl}}\right)& \mbox{for}& H\geq H_E\\
-M_{pl}\sqrt{\frac{2}{3}}\ln\left(\frac{H-H_f}{H_E-H_f}\right)+\varphi_E &\mbox{for}& H \leq H_E,
\end{array}\right.
\end{eqnarray}
where $\varphi_E=M_{pl}\sqrt{\frac{2}{k}}\ln\left(\frac{M_{pl}-H_E}{M_{pl}}\right)\approx -\sqrt{\frac{2}{k}}H_E$.

\

The potential is given by $V(H)=3H^2M_{pl}^2+\dot{H}M_{pl}^2$. Hence,
\begin{eqnarray}
V(\varphi)=\left\{\begin{array}{ccc}
M_{pl}^4\left[3\left(1-e^{\frac{\varphi}{M_{pl}}\sqrt{\frac{k}{2}}}\right)^2-ke^{\frac{\varphi}{M_{pl}}\sqrt{2k}}\right] & \mbox{for}& \varphi\leq \varphi_E\\
3M_{pl}^2H_f^2\left[1+2\left(\frac{H_E}{H_f}-1\right)e^{-\sqrt{\frac{3}{2}}\frac{\varphi-\varphi_E}{M_{pl}}}\right] &\mbox{for}& \varphi\geq \varphi_E.
\end{array}\right. \label{potential}
\end{eqnarray}

\


Now, we aim to analyze the stability of the solution to our model. Using the potential found in Eq. \eqref{potential}, we will study the dynamics of the equation:
\begin{eqnarray}
\ddot{\varphi}+3H(\varphi,\dot{\varphi})\dot{\varphi}+V_{\varphi}=0,
\end{eqnarray}
where $H(\varphi,\dot{\varphi})=\sqrt{\frac{1}{3}\left(\frac{\dot{\varphi}^2}{2}+V(\varphi) \right)}$.

\

\begin{figure}[H]
\begin{center}
\includegraphics[scale=0.33]{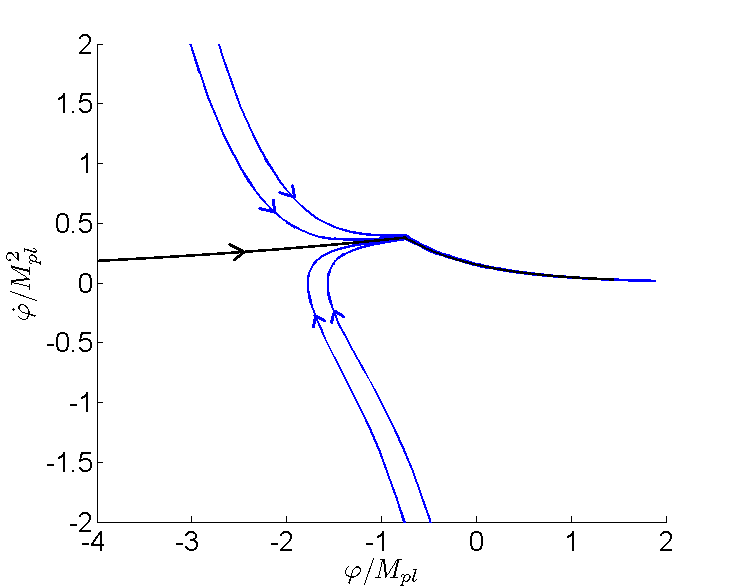}
\includegraphics[scale=0.42]{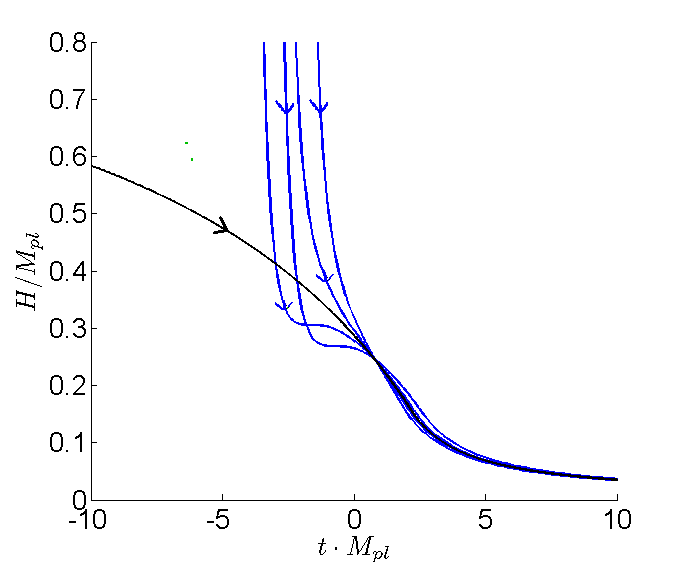}
\end{center}
\caption{Phase portrait in the plane $(\varphi/M_{pl},\dot{\varphi}/M_{pl}^2)$ (left) for some orbits, with the analytical one represented in black. Evolution of $H/M_{pl}$ in fuction of the time $t\times M_{pl}$ (right) for the same orbits represented in the phase portrait.}
\label{fig:dynamics}
\end{figure}

The numerical results are presented in Figure \ref{fig:dynamics}. For clarity in the understanding of the dynamical system, we have used a value of $k$ which differs in various orders of magnitude from the one in our model. Though, the behaviour is exactly the same as with the real values. Therefore we observe that the analytical orbit is the only one with 2 de Sitter points. All the other orbits start with an infinite energy at $t\to-\infty$, they experience a phase transition at $\varphi=\varphi_E$ and then asymptotically approach the analytical orbit for $t\to\infty$, reaching at $t\to\infty$ the de Sitter point $H=H_f$.

\

An important remark is in order: From the phase space portrait we can see that there is only one non-stable non-singular solution, i.e., if one takes initial conditions near the analytical solution one obtains a past singular orbit. This is exactly the same that happens in the Starobinsky model \cite{starobinsky} (see Figure 3 of \cite{aho} for details).

\subsection{Reheating constraints}

Before studying the reheating constraints in our model, we are going to carry out some useful approximations that will help us in our further calculations. Since  $ k\cong\frac{10^9H_*^4}{16\pi^2M_{pl}^4}$,  we easily obtain $H_E\cong \sqrt{\frac{k}{3}} M_{pl}\cong  \frac{10^4}{4\pi}\sqrt{\frac{10}{3}} \frac{H_*^2}{M_{pl}^2}M_{pl}$. On the other hand, it is also fulfilled that $0\cong\alpha M_{pl}^2-2H_*^2$ and, thus, we obtain all the following approximative expressions
\begin{eqnarray}
\left. \begin{array}{cc}
 k\cong 10^{-9} \pi^2(1-n_s)^2 ,\ \ \ \ \ \ \alpha\cong 8\times10^{-9}\pi^2(1-n_s) \\ H_{*}\cong  6\times 10^{-5}\pi \sqrt{1-n_s} M_{pl}, \ \
  \ \ \ \ \ \ H_{end} \cong 3\times 10^{-5}\pi(1-n_s)M_{pl},\ \ \ \ \ \ \ H_E\cong 2\times 10^{-5}\pi(1-n_s) M_{pl} ,\end{array}\right.
\end{eqnarray}
and, hence,  $r=16\epsilon_*\cong 16k\frac{M_{pl}^2}{H_*^2}\cong 4(1-n_s)$.

Moreover, the number of e-folds is
\begin{eqnarray}
N\cong \frac{1}{k}\left(\ln\left( \frac{1-\sqrt{\alpha/2}}{1-\sqrt{k}} \right)+ \frac{1}{1-\sqrt{\alpha/2}}-\frac{1}{1-\sqrt{k}}\right)\cong \frac{\alpha}{4k}-\frac{1}{2}\cong \frac{2}{1-n_s}-\frac{1}{2},
\end{eqnarray}
but using equation \eqref{eq13} we also obtain
\begin{eqnarray}
N\cong 71.1+\frac{1}{6}\ln(1-n_s)-\frac{1}{3}\ln\left( \frac{g_R^{\frac{1}{4}}T_R}{\mbox{GeV}} \right).
\end{eqnarray}

\

Equaling both quantities one obtains the equation
\begin{eqnarray}
Y+\frac{1}{2}\ln Y=217.5-\ln\left( \frac{g_R^{\frac{1}{4}}T_R}{\mbox{GeV}} \right), \label{eq22}
\end{eqnarray}
where we have introduced the notation $Y=\frac{6}{1-n_s}$.

\

\begin{figure}[H]
\begin{center}
\includegraphics[height=45mm]{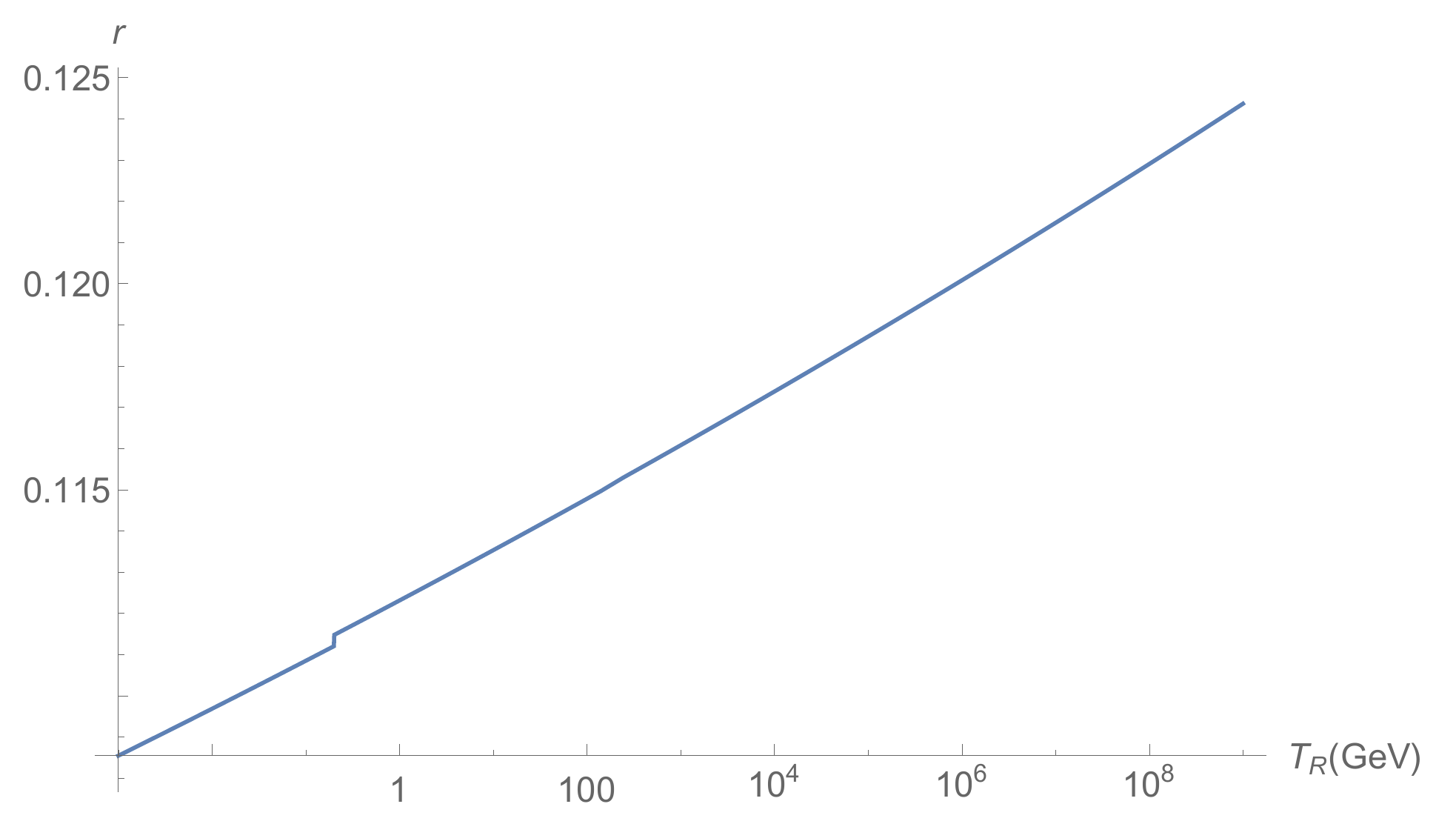}
\includegraphics[height=45mm]{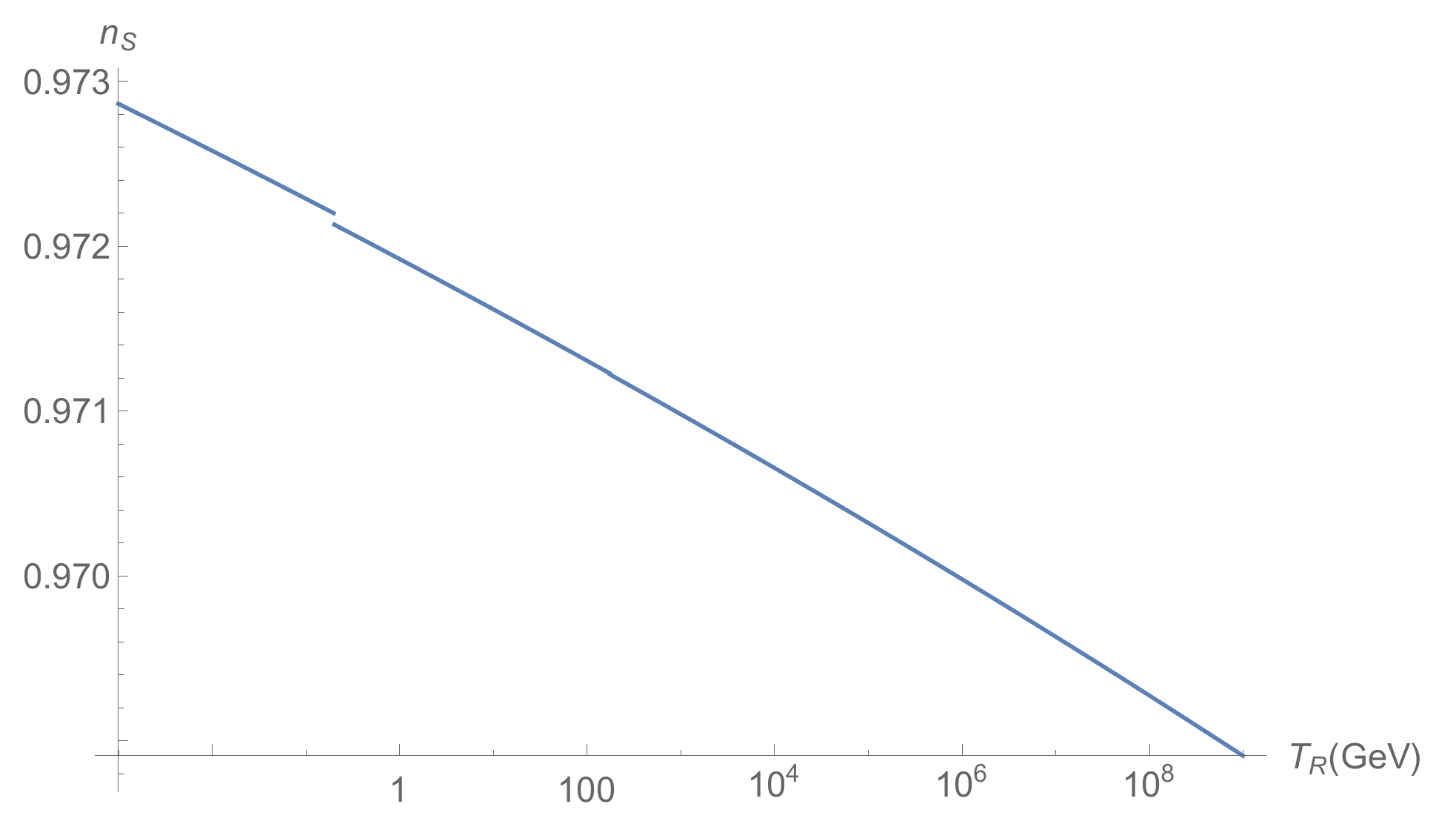}
\end{center}
\caption{Evolution of the tensor/scalar ratio $r$ (left) and the spectral index $n_s$ (right) versus the reheating temperature $T_R$.}
\label{fig:rvsT}
\end{figure}

Hence, from this equation we obtain the following relation between the cosmological parameters ($n_s$,$r$) and $T_R$, showing that the suitable constraints will be verified for $1 \text{ MeV}\leq T_R\leq 10^5$ GeV, as we can see in Figure \ref{fig:rvsT}.

\

Now, let us consider the gravitational production of $\chi$-particles. Firstly, we treat the case when the produced particles are very massive and conformally coupled with gravity. Since a classical picture of the universe is only possible at energy densities less than the Planck's one, it seems natural to choose that at this scale
the quantum field $\chi$ is in the vacuum. In fact, we will choose as a vacuum state the adiabatic one, defined by the modes \cite{kaya}
\begin{eqnarray}
\chi_k=\frac{1}{\sqrt{2\Omega_k}}e^{-i\int \Omega_k d\eta},
\end{eqnarray}
where $\Omega_k$ satisfies the equation
\begin{eqnarray}
\Omega_k^2= (k^2+a^2m^2)+\frac{3\Omega_k'^2}{4\Omega_k^2}-\frac{\Omega_k''}{2\Omega_k}.\end{eqnarray}.

Then, to obtain an approximate expression of $\Omega_k$, one can use the WKB approximation, which holds when $m\geq H$. For this reason, since at the Planck epoch $H\sim M_{pl}$, in order to obtain the WKB solution one has to assume that the mass of the field, namely $m$, satisfies $m\geq M_{pl}$. However, since for $m> \sqrt{4\pi} M_{pl}$ the produced particles are micro Black Holes \cite{FKL}, whose thermodynamical description is unknown \cite{Helfer}, one has to choose $m\sim M_{pl}$ so as to avoid this Black Holes production.

\

The reheating temperature caused by the decay of these particles into lighter ones, with a thermalization rate - the same used in \cite{pv,Allahverdi}-  equal to $\Gamma=\frac{\beta^2}{m}\rho_{\chi}^{1/2}$ 
with $\beta^2\sim 10^{-3}$ \cite{spokoiny}, is of the order (see for details \cite{inflation2,has}) 
\begin{eqnarray}
T_R\sim 10^{-1}\left(\frac{H_E}{M_{pl}} \right)^2\left(\frac{H_E}{m}\right)M_{pl}\sim 2\times 10^{-13}(1-n_s)^3\frac{M_{pl}^2}{m}\sim 
2\times 10^{-13}(1-n_s)^3 M_{pl}\end{eqnarray}
and, inserting this expression in \eqref{eq22}, one obtains
\begin{eqnarray}
Y-\frac{5}{2}\ln Y \cong 197.9,
\end{eqnarray}
which leads to a spectral index $n_s\cong 0.9716$, a tensor/scalar ratio $r\cong 0.1136$ and a reheating temperature $T_R\sim 11$ GeV. { Since this model does not come from supergravity, we do not need to take into account the gravitino and moduli fields problems. Hence, these results lead to a viable model}.

\

On the other hand, when considering massless particles nearly conformally coupled with gravity, the reheating temperature becomes \cite{pv,inflation1}
\begin{eqnarray}
T_R\sim\mathcal{N}^{3/4}\left|\xi-\frac{1}{6}\right|^{3/2}\frac{H_E^2}{M_{pl}^2}M_{pl}\sim 3\times 10^{-9}\left|\xi-\frac{1}{6}\right|^{3/2}(1-n_s)^2M_{pl} \label{eq24}
\end{eqnarray}
 with $\xi$ the coupling constant,
$\mathcal{N}=\frac{1}{8\pi^2}\int_0^{\infty}s|g(s)|^2 ds$, being $g(s)=\frac{1}{H_E^2a_E^2}\int_{-\infty}^{\infty}e^{-2is\tau}a^2\left(\frac{\tau}{H_E}\right)R\left(\frac{\tau}{H_E}\right)\frac{d\tau}{H_E}$, 
where $R$ is the scalar curvature. It has been numerically computed that $\mathcal{N}\sim 1$. So as to verify the bounds $1\text{ MeV}\leq T_R\leq 10^5$ GeV (coming from the restriction $r\leq 0.12$ in our model), the coupling constant $\xi$ must satisfy
\begin{eqnarray}
3\times 10^{-7}\lesssim\left|\xi-\frac{1}{6}\right| \lesssim 6\times 10^{-2}.
\end{eqnarray}

If we consider now massless particles far from the conformal coupling with gravity, { using the results of \cite{dv,Giovannini} which we will discuss in the appendix}, the reheating temperature was calculated in \cite{has}, leading to
\begin{eqnarray}
T_R\sim 3\times 10^{-2}\frac{H_E^2}{M_{pl}^2}M_{pl}\sim 10^{-10}(1-n_s)^2M_{pl}\sim 2\times 10^5 \text{ GeV}, \label{eq26}
\end{eqnarray}

In this case combining equations \eqref{eq22} and \eqref{eq26}, one obtains
\begin{eqnarray}
Y-\frac{3}{2}\ln Y \cong 193.5,
\end{eqnarray}
whose solution is $n_s=0.970$, corresponding to $r=0.119$, coinciding thus with the limit where our model starts being valid. Hence, our model supports the presence of these particles. 

\section{Simple quintessential inflation models}

In this section, we are going to proceed in an inverse way from the one we have just done. We are going to take some simple and well-known potentials, some of them with a similar form from the one already studied, and we are going to verify whether we can adjust the parameters so as to fulfill all the corresponding constraints for the cosmological parameters. To simplify,
we will consider some well-known positive inflationary potentials that vanish at some value of the scalar field, namely $\varphi_E$, and we will extend it to zero for the other values of the field. Moreover, we introduce a cosmological constant $\Lambda\sim 3H_0^2$ (being $H_0$ the current Hubble cosmological constant) to ensure the current cosmic acceleration.

\subsection{Exponential SUSY Inflation (ESI)}

The first potential we are going to study is  an Exponential SUSY Inflation (ESI) style potential  \cite{Obukhov, DT},
\begin{eqnarray}
V(\varphi)=\left\{ \begin{array}{ll} \lambda M_{pl}^4 (1-e^{\frac{\varphi}{M_{pl}}}) & \mbox{$\varphi<0$} \\ 0 & \mbox{$\varphi\geq 0$},  \end{array} \right.
\end{eqnarray}
being $\lambda$ a dimensionless positive parameter. By using the following approximate expressions of the slow-roll parameters as a function of the potential,
\begin{eqnarray}
\epsilon\approx\frac{M_{pl}^2}{2}\left(\frac{V_{\varphi}}{V}\right)^2 \ \ \ \eta\approx M_{pl}^2\frac{V_{\varphi\varphi}}{V},
\end{eqnarray}
we obtain $\epsilon_*= \frac{s_*^2}{2}$ and $\eta_*=-s_*$, where $s_*=\frac{e^{\frac{\varphi_*}{M_{pl}}}}{1-e^{\frac{\varphi_*}{M_{pl}}}}\cong e^{\frac{\varphi_*}{M_{pl}}}$. Hence, we can compute the associated inflationary parameters, analogously as we did in Section $3$:
\begin{eqnarray}
n_s-1=-s_*\left(\frac{3}{2}s_*+2\right) \ \ \ \ \alpha_s=-\frac{\sqrt{2\epsilon_*}}{1-\epsilon_*}\frac{3s_*+2}{4\sinh^2\left(\frac{\varphi_*}{2M_{pl}} \right)} \ \ \ \ r=16\epsilon_*, \label{eqpot1}
\end{eqnarray}
where we have used for the calculation of $\alpha_s$ that $\dot{H}=-\frac{\dot{\varphi}^2}{2M_{pl}^2}$. It is also straightforward to calculate the power spectrum:
\begin{eqnarray}
P\approx \frac{H_*^2}{8\pi^2\epsilon_*M_{pl}^2}\approx \frac{V(\varphi_*)}{24\pi^2M_{pl}^4\epsilon_*}=\frac{\lambda (1-e^{\frac{\varphi_*}{M_{pl}}})}{12\pi^2 s_*^2},
\end{eqnarray}
where we have used that $V(\varphi_*)\approx 3M_{pl}^2H_*^2$. We will guarantee that the value $P\sim 2\times 10^{-9}$ is verified in our model by taking $\lambda=\frac{24\times 10^{-9}\pi^2s_*^2}{1-e^{\frac{\varphi_*}{M_{pl}}}}$ . Finally, regarding the number of e-folds,
\begin{eqnarray}
N=\int_{t_*}^{t_{end}}Hdt=\frac{1}{M_{pl}}\int_{\varphi_*}^{\varphi_{end}}\frac{1}{\sqrt{2\epsilon}}d\varphi=e^{-\frac{\varphi_*}{M_{pl}}}-\frac{1+\sqrt{2}}{\sqrt{2}}+\frac{\varphi_*}{M_{pl}}-\ln\left(\frac{\sqrt{2}}{1+\sqrt{2}} \right).
\end{eqnarray}

So as to fit our model with the recent observational values, we will initially take, as before, the spectral index obtained by Planck2015 data, $n_s=0.968\pm 0.006$ and we will find the corresponding value $\varphi_*$ from the first equation in \eqref{eqpot1}, which is $-4.37\leq \frac{\varphi_*}{M_{pl}}\leq -3.99$. Therefore, the other cosmological parameters are $0.0013 \leq r\leq 0.0028$ and $-7\times 10^{-4}\leq\alpha_s\leq -3\times 10^{-4}$. And, finally, the corresponding number of e-folds is $49\lesssim N\lesssim 73$.

\

\begin{figure}[H]
\begin{center}
\includegraphics[scale=0.45]{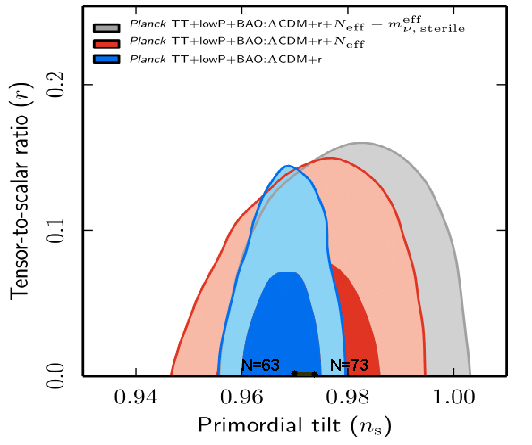}
\end{center}
\caption{Marginalized joint confidence contours for $(n_s, r)$, at the 68\% and 95\% CL, without the presence of running of the spectral indices. We have drawn the curve from $N=63$ to $73$ e-folds. (Figure courtesy of the Planck2015 Collaboration).}
\label{fig:planck2}
\end{figure}

Hence, all the values of $r$ and $\alpha_s$ fit to our restrictions and, regarding the number of e-folds, in order to have $63\lesssim N\lesssim 73$ we need to take the range of $0.970\leq n_s\leq 0.974$ for the spectral index, obtaining thus $0.0013<r<0.0017$ and $-5\times 10^{-4}<\alpha_s<-3\times 10^{-4}$. Moreover, the theoretical values provided by our model enters in the marginalized $1\sigma$ C.L. contour in the plane $(n_s,r)$ as one can see in Figure \ref{fig:planck2}.

Since the approximative expression of the cosmological parameters is only valid during the inflation process, we cannot directly calculate the Hubble constant at the transition, namely $H_E$. As we can verify in the former model where we had the analytical form of the dynamical system, it is of the same order as at the end of the inflation ($H_{end}$). Since $\epsilon_{end}=1$ and
$\varphi_{end}=\ln\left(\frac{\sqrt{2}}{1+\sqrt{2}}\right)M_{pl}$ and using that
$\epsilon_{end}=1 \Longrightarrow  \dot{H}_{end}=-H_{end}^2$ 
and $V=3H^2M_{pl}^2+\dot{H}M_{pl}^2$,
one can conclude that
$H_{end}=\sqrt{\frac{V(\varphi_{end})}{2M_{pl}^2} }=M_{pl}\sqrt{\frac{\lambda}{2(1+\sqrt{2})}}$, meaning that for our model one has $H_{end}\sim 3\times 10^{-6}M_{pl}$.

\

\begin{figure}[H]
\begin{center}
\includegraphics[height=45mm]{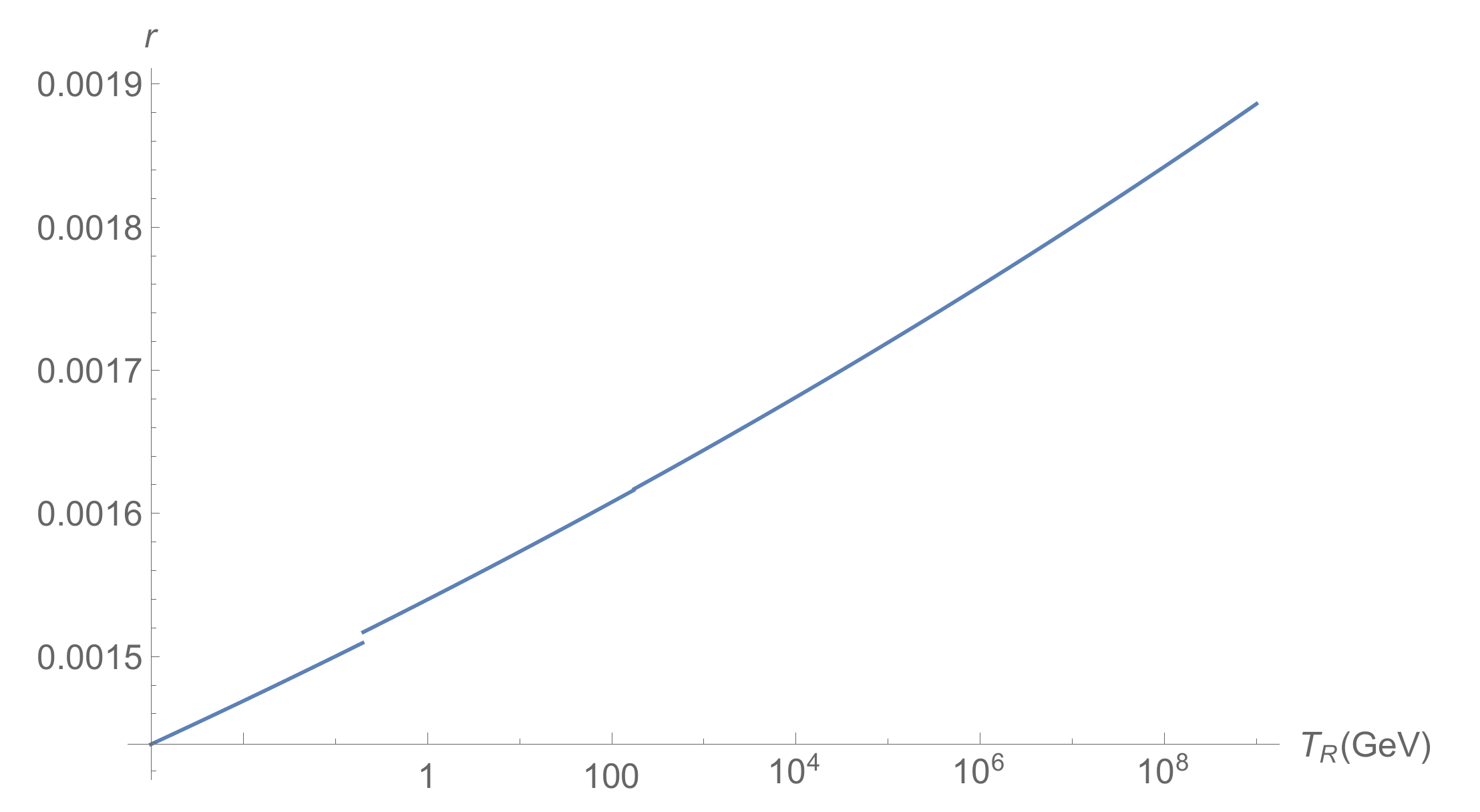}
\includegraphics[height=45mm]{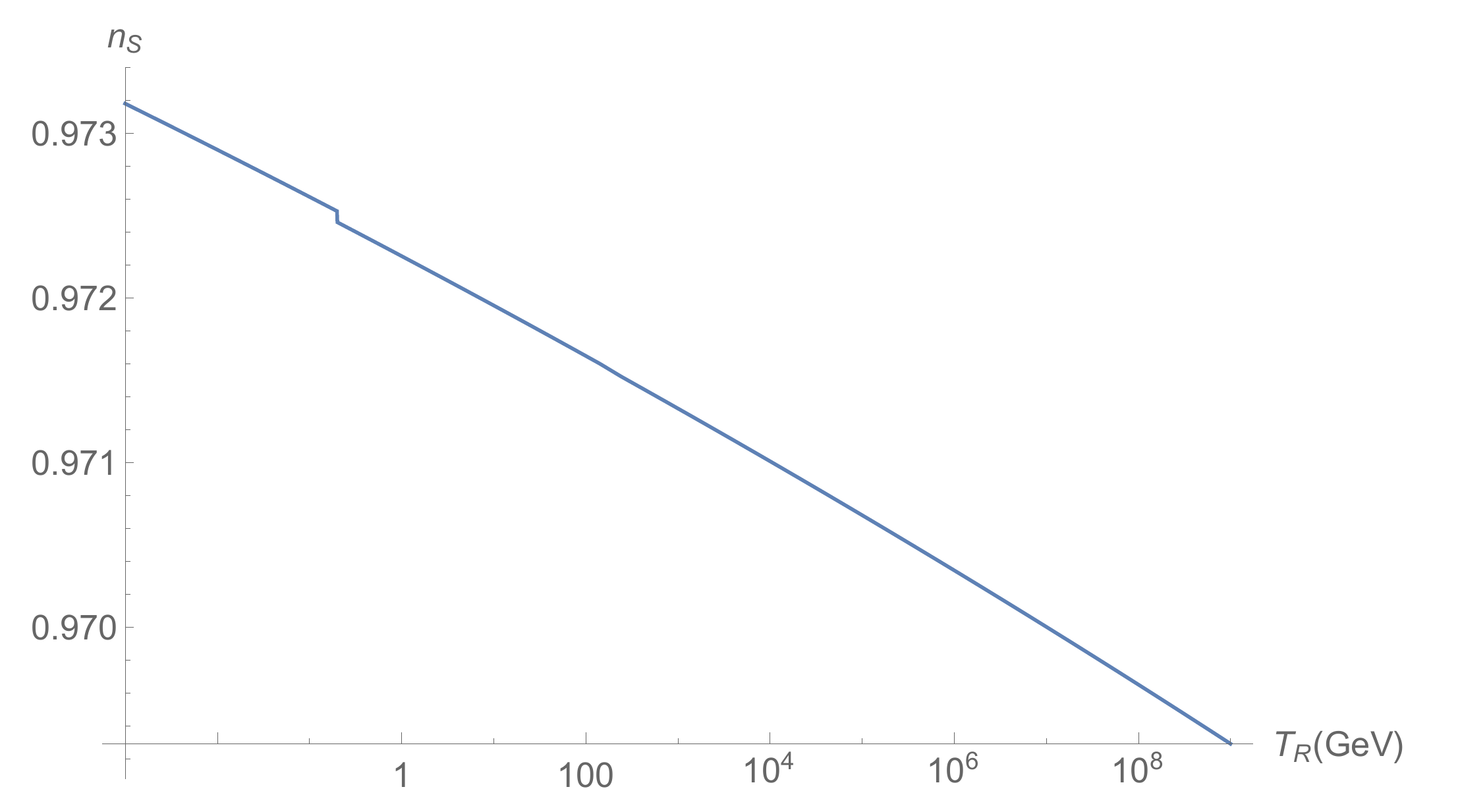}
\end{center}
\caption{Evolution of the tensor/scalar ratio $r$ (left) and the spectral index $n_s$ (right) versus the reheating temperature $T_R$.}
\label{fig:rvsT2}
\end{figure}

So as to study the reheating constraints for this model, we are going to proceed as in the former case, i.e, with the approximative expressions $r\approx 2(1-n_s)^2$, $H_E\sim H_{end}\approx \frac{\sqrt{\lambda}}{2}M_{pl}\approx\frac{\pi(1-n_s)\sqrt{3P}}{2}\approx 10^{-4}(1-n_s)M_{pl}$ and $N\approx \frac{2}{1-n_s}+\ln\left(\frac{1-n_s}{2} \right)$.  Now we can not reproduce the same calculation as we did for the former case so as to obtain Equation \eqref{eq13}, because we do not have an analytic expression to perform the calculations given in (\ref{eq 12}), but since
this value is negligible  we can continue using it.
 Therefore, by combining equation \eqref{eq13} and the just obtained approximate value of $N$, we have that
\begin{eqnarray}
Y-\ln Y=212.0-\ln\left(\frac{g_R^{1/4}T_R}{\text{GeV}} \right), \label{eq33}
\end{eqnarray}
again using that $Y=\frac{6}{1-n_s}$. Hence, from solving this equation, one can see in Figure \ref{fig:rvsT2} that the constraints on $r$ and $n_s$ are fulfilled, as we had already pointed out.

Now, with regards to the case when the produced particles are very massive and conformally coupled with gravity (again choosing $m\sim M_{pl}$ for the same argument as in previous section), we 
{have to proceed as follows:

The square modulus of the $\beta$-Bogoliubov coefficient is given by the expression \cite{inflation2}
\begin{eqnarray}\label{beta}
|\beta_k|^2\cong \frac{m^4 a_E^{10}(\ddot{H}_E^+-\ddot{H}_E^-)^2}{256(k^2+m^2a_E^2)^5},
\end{eqnarray}
where $a_E$ is the value of the scale factor at the transition phase, and $\ddot{H}_E^{\pm}$ are the value of the second derivative of the Hubble parameter after and before the phase transition.
Since we do not have an analytic expression of the Hubble parameter, to obtain its second derivative we use the equation $\ddot{H}=-\frac{\dot{\varphi}\ddot{\varphi}}{M_{pl}^2}$. From the
conservation equation we deduce that $|\ddot{\varphi}_E^+-\ddot{\varphi}_E^-|=|V_{\varphi}(0^-)|=\lambda M_{pl}^3$. On the other hand,  
at the transition time all the energy density is kinetic, which means that $\dot{\varphi}_E=\sqrt{6}{H}_EM_{pl}
\cong \sqrt{6}{H}_{end}M_{pl}=  \sqrt{\frac{3\lambda}{2}} M_{pl}^2$, and thus,
$
(\ddot{H}_E^+-\ddot{H}_E^-)^2\cong \frac{3\lambda^3}{2}M_{pl}^6\cong 32 H_{end}^6.
$

A  simple integration leads to an energy density for the produced particles of
$\rho_{\chi}\cong \frac{5H_{end}^6}{16^3\pi m^2}$, and following the calculations made in \cite{inflation2}
one gets the following reheating temperature
\begin{eqnarray}
T_R\sim \frac{1}{\sqrt{3}}\left(\frac{5}{16\pi}\right)^{5/8}\left(\frac{H_{end}}{M_{pl}}\right)^2\frac{H_{end}}{m} M_{pl}\sim
10^{-1}\left( \frac{H_{end}}{M_{pl}}\right)^2\left(\frac{H_{end}}{m}\right) M_{pl} \sim 10^{-13}(1-n_s)^3M_{pl}\end{eqnarray}
which leads, by combination with \eqref{eq33}, to
\begin{eqnarray}
Y-4\ln Y\cong 193.1,
\end{eqnarray}
obtaining that $n_s\cong 0.9720$, $r\cong 1.56\times 10^{-3}$ and $T_R\sim 5$ GeV. { This result again satisfies the gravitino-overproduction problem but not the moduli fields one.}

\

In order to study massless particles nearly conformally coupled with gravity, we will use the same formula as used in \eqref{eq24} and, since we do not have the analytical expression of the Hubble constant throughout all the time, we will assume that ${\mathcal N\sim 1}$ as in the other case. Hence, the reheating temperature is
\begin{eqnarray}
T_R\sim 10^{-8}\left|\xi-\frac{1}{6} \right|^{3/2}(1-n_s)^2 M_{pl}
\end{eqnarray}
obtaining, thus, that for the bounds of $T_R$ coming from the nucleosynthesis it should be verified that  $10^{-7}\lesssim\left|\xi-\frac{1}{6}\right|\lesssim 16$. This only gives us a restriction for the lower bound, since given that we have considered the particles to be nearly conformally coupled with gravity $\left|\xi-\frac{1}{6}\right|\lesssim 10^{-1}${, which adds the constraint that $T_R\leq 7\times 10^5$ GeV. By considering a reheating temperature in the MeV scale, one would obtain that $10^{-7}\lesssim\left|\xi-\frac{1}{6} \right|\lesssim 10^{-5} $}. Finally, regarding massless particles far from the conformal coupling with gravity, the reheating temperature becomes
\begin{eqnarray}
T_R\sim 3\times 10^{-2}\frac{H_E^2}{M_{pl}^2}M_{pl}\sim 3\times 10^{-10}(1-n_s)^2M_{pl}\sim 6\times 10^5 \text{ GeV},
\end{eqnarray}
which leads us, using also Equation \eqref{eq33}, that
\begin{eqnarray}
Y-3\ln Y\cong 186.9,
\end{eqnarray}
whose solution is $n_s=0.970$, corresponding to $r=0.0018$, which now clearly falls within the bounds imposed in our model so as to satisfy the observational data{, but again fails to fulfill the bound $T_R\leq 1$ GeV suggested in the introduction.}

\

Finally we are going to briefly study the dynamics of the equation $\ddot{\varphi}+3H(\varphi,\dot{\varphi})\dot{\varphi}+V_{\varphi}=0$. In Figure \ref{fig:dynamicspot1} we have represented in blue some orbits which are solution of these dynamical system, observing that, as in the former case, they start with infinite energy at $t\to-\infty$, they experience a phase transition at $\varphi=0$ and then evolve towards a de Sitter phase for $t\to\infty$. Moreover, we have drawn in black the slow-roll solution $\dot{\varphi}=-\frac{V_{\varphi}}{3H}=-\frac{V_{\varphi} M_{pl}}{\sqrt{3V(\varphi)}}=\sqrt{\frac{2\epsilon(\varphi) V(\varphi)}{3}}$.

\begin{figure}[H]
\begin{center}
\includegraphics[scale=0.40]{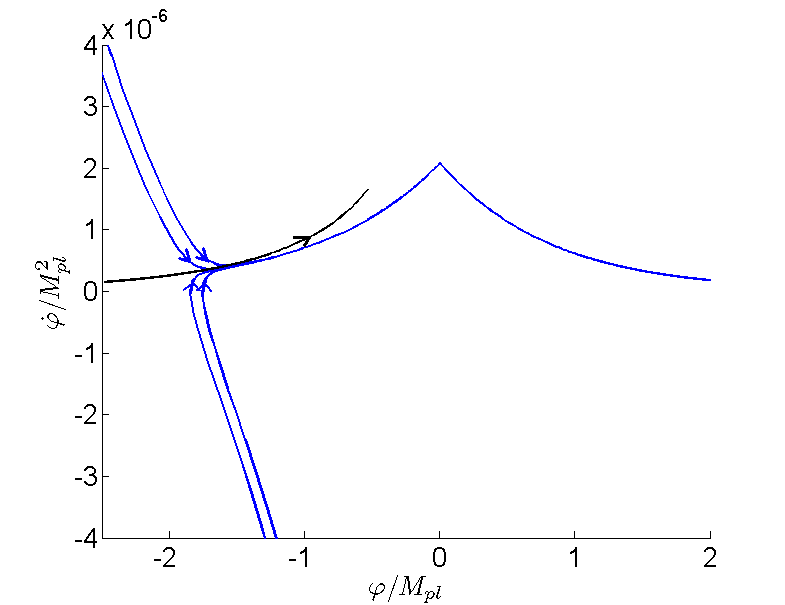}
\includegraphics[scale=0.47]{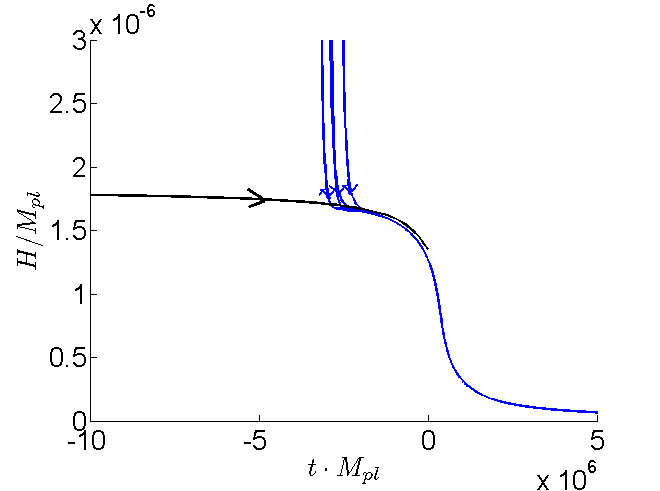}
\end{center}
\caption{Phase portrait in the plane $(\varphi/M_{pl},\dot{\varphi}/M_{pl}^2)$ (left) for some orbits, with the slow-roll one represented in black. Evolution of $H/M_{pl}$ in fuction of the time $t\times M_{pl}$ (right) for the same orbits represented in the phase portrait.}
\label{fig:dynamicspot1}
\end{figure}

We observe that there is a long period in inflation where orbits coincide with the slow-roll one. Furthermore, as happened with the analytical solution in the former case we studied, the slow roll solution is a limit orbit that starts with a de Sitter phase at $t\to-\infty$, since all the other orbits start with infinite energy.

\subsection{Higgs Inflation (HI)}

Another potential that could work would be the following Higgs Inflation (HI) style potential in the Einstein Frame \cite{enciclopedia}.
\begin{eqnarray}
V(\varphi)=\left\{ \begin{array}{ll} \lambda M_{pl}^4 \left(1-e^{\frac{\varphi}{M_{pl}}}\right)^2, & \mbox{$\varphi<0$} \\ 0, & \mbox{$\varphi\geq 0$}  \end{array} \right.
\end{eqnarray}
being $\lambda$ a dimensionless positive parameter. In this case, the slow-roll parameters are $\epsilon_*=2s_*^2$ and $\eta_*=-s_*(1-s_*)$, being $s_*$ the same as above. In this case, the running, power spectrum and number of e-folds are
\begin{eqnarray}
\left. \begin{array}{cc} \alpha_s=-\frac{\sqrt{2\epsilon_*}}{1-\epsilon_*}2e^{\frac{\varphi_*}{M_{pl}}}\left(1+10(1+s_*)^2 \right) \ \ \ \ \ \ \  \ \ \ \ \ \  P\approx \frac{\lambda\left(1-e^{\frac{\varphi_*}{M_{pl}}} \right)^2}{12\pi^2s_*^2} \\
N=\int_{t_*}^{t_{end}}Hdt=\frac{1}{M_{pl}}\int_{\varphi_*}^{\varphi_{end}}\frac{1}{\sqrt{2\epsilon}}d\varphi=\frac{e^{-\frac{\varphi_*}{M_{pl}}}-(1+\sqrt{2})}{2}+\frac{\varphi_*}{2M_{pl}}-\frac{1}{2}\ln\left(\frac{1}{1+\sqrt{2}} \right). \end{array}\right.
\end{eqnarray}

However, in this model now we have $N\sim \frac{1}{s_*}\cong \frac{1}{1-n_s}$, which is bounded by $50$ for the allowed values of the spectral index. Thus, the potential is not a viable quintessential inflation model because it leads to an insufficient number of e-folds.

\subsection{Power Law Inflation (PLI)}

Now, we are going to study a Power Law Inflation (PLI) \cite{Sahni} potential, adapted to quintessence
\begin{eqnarray}
V(\varphi)=\left\{ \begin{array}{ll} \lambda M_{pl}^4 \left(\frac{\varphi}{M_{pl}}\right)^{2n}, & \mbox{$\varphi<0$} \\ 0, & \mbox{$\varphi\geq 0$}  \end{array} \right.
\end{eqnarray}

With the same procedure as in the former cases, we obtain $\epsilon_*=2n^2\left(\frac{M_{pl}}{\varphi_*}\right)^2$ and $\eta=2n(2n-1)\left(\frac{M_{pl}}{\varphi}\right)^2$. So, $\varphi_*=-M_{pl}\sqrt{\frac{4n(n+1)}{1-n_s}}$ and $N=\frac{n+1}{1-n_s}-\frac{n}{2}$. Hence, we can easily verify that, so as to obtain a number of e-folds $63\leq N\leq 73$ with a spectral index $n_s=0.968\pm 0.006$, we need that $0.65<n<1.35$. 

\

Therefore, taking $n=1$ is a good choice in order to match our model with the observational results. Thus, if we consider the range of spectral index $0.9685\leq n_s\leq 0.973$, we obtain the desired number of e-folds. Regarding the ratio of tensor to scalar perturbations $r=16\epsilon_*=4(1-n_s)$, the constraint $r\leq 0.12$ is verified for $n_s\geq 0.97$. And so does the running, which becomes $\alpha_s=-\frac{8\sqrt{2\epsilon_*}}{1-\epsilon_*}\left(\frac{M_{pl}}{\varphi_*}\right)^2\approx -16\left|\frac{M_{pl}}{\varphi_*}\right|^3=-\frac{(1-n_s)^{3/2}}{\sqrt{2}}$, resulting that $-3.7\times 10^{-3}\leq \alpha_s\leq -3.1\times 10^{-3}$ for $0.97\leq n_s\leq 0.973$. Finally, the power spectrum has the following expression

\begin{eqnarray}
P\approx \frac{\lambda}{48\pi^2}\left(\frac{\varphi_*}{M_{pl}}\right)^4=\frac{4\lambda}{3\pi^2(1-n_s)^2}
\end{eqnarray}
and, thus, since $P\sim 2\times 10^{-9}$, we can determine $\lambda$, namely $\lambda\sim 10^{-11}$.

\

\textbf{Remark:} We can see that the power law potential for $n=2$ is equivalent during the inflation, i.e. for $\varphi\ll -|\varphi_E|$ to the Double Well Inflation (DWI) \cite{Goldstone} potential, namely

\begin{eqnarray}
V(\varphi)=\left\{ \begin{array}{ll} \lambda (\varphi^2-\varphi_E^2)^2, & \mbox{$\varphi<\varphi_E$} \\ 0, & \mbox{$\varphi\geq \varphi_E$.}  \end{array} \right.
\end{eqnarray}

Therefore, we have seen that the DWI is not a valid model that matches with the corresponding observational results.

\

Now, we are going to proceed analogously as for the potential previously studied. Therefore, we approximate the Hubble constant at the transition point by $H_E\sim H_{end}=M_{pl}\sqrt{\lambda}\approx 10^{-4}(1-n_s)M_{pl}$. Now, combining Equation \eqref{eq13} and the number of e-folds for this particular potencial, we obtain that

\begin{eqnarray}
Y+\frac{1}{2}\ln Y =213.5-\ln\left(\frac{g_R^{1/4}T_R}{\text{GeV}}\right), \label{eq47}
\end{eqnarray}
where, as before, $Y=\frac{6}{1-n_s}$. Thus, as we can see in Figure \ref{fig:rvsT3}, the bounds coming from the observational values used throughout all this paper are satisfied for $1\text{ MeV}\leq T_R\leq 10^4$ GeV (corresponding to $0.970\leq n_s\leq 0.973$).
\begin{figure}[H]
\begin{center}
\includegraphics[height=45mm]{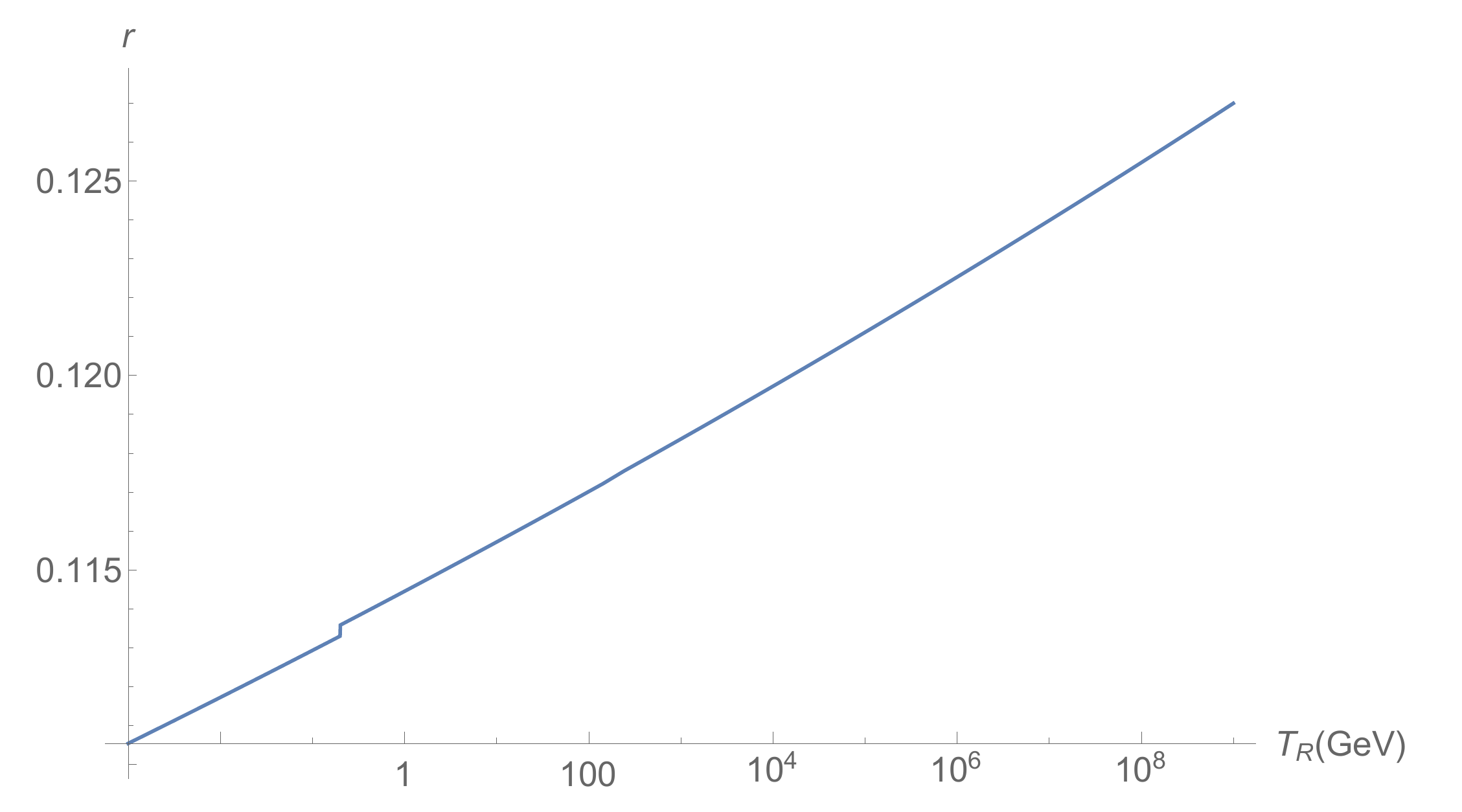}
\includegraphics[height=45mm]{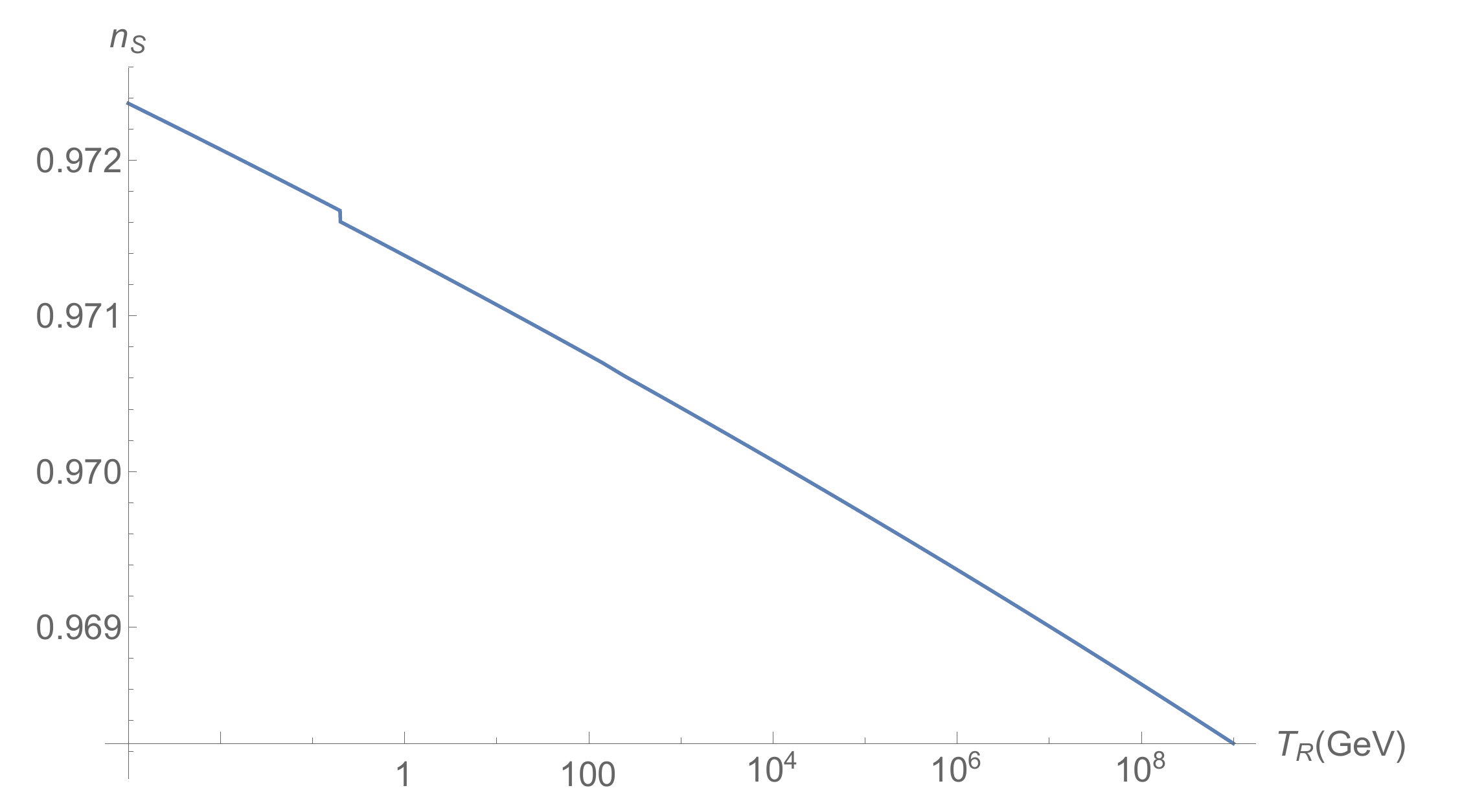}
\end{center}
\caption{Evolution of the tensor/scalar ratio $r$ (left) and the spectral index $n_s$ (right) versus the reheating temperature $T_R$.}
\label{fig:rvsT3}
\end{figure}

\


If we start considering the production of massless particles nearly conformally coupled to gravity, then the reheating temperature becomes
\begin{eqnarray}
T_R\sim \left|\xi-\frac{1}{6} \right|^{3/2}{\mathcal N}^{3/4}\frac{H_E^2}{M_{pl}}\sim 10^{-8}\left|\xi-\frac{1}{6}\right|^{3/2} (1-n_s)^2 M_{pl}.
\end{eqnarray}

Hence, for $1\text{ MeV} \leq T_R\leq 10^4$ GeV, we obtain that $10^{-7}\lesssim \left|\xi-\frac{1}{6}\right|\lesssim 5\times 10^{-3}$. On the other hand, in the case of heavy massive particles ($m\sim M_{pl}$), 
since the first derivative of the potential is continuous at the transition phase, one has to use, for the $\beta$-Bogoliubov coefficient the expression given in \cite{inflation1},
\begin{eqnarray}
|\beta_k|^2\cong \frac{m^4 a_E^{12}(\dddot{H}_E^+-\dddot{H}_E^-)^2}{1024(k^2+m^2a_E^2)^6}.
\end{eqnarray}

To obtain the third derivative of the Hubble parameter first of all we use the formula $\dddot{H}=-\frac{\ddot{\varphi}^2+\dot{\varphi}\dddot{\varphi}}{M_{pl}^2}$. Using the conservation equation one gets
$|\dddot{\varphi}_E^+-\dddot{\varphi}_E^-|=|V_{\varphi\varphi}(0^-)\dot{\varphi}_E|\cong 2\sqrt{6}\lambda H_{end} M_{pl}^3$, where we have used that at the transition time all energy density is kinetic.
Then we have 
$(\dddot{H}_E^+-\dddot{H}_E^-)^2\cong 144\lambda^2 H_{end}^4M_{pl}^4=144 H_{end}^8$, where we have used that $H_{end}=\sqrt{\lambda}M_{pl}$. Therefore, the energy density of produced particles is equal to $\rho_{\chi}\cong \frac{10^{-3}}{\pi}\frac{H_{end}^8}{m^4}$, thus,  following step by step the calculations made in \cite{inflation2} and, using the thermalization rate
introduced in Section $2.3$, one gets the following reheating temperature
\begin{eqnarray}
T_R\sim5 \times 10^{-3}\left(\frac{H_{end}}{M_{pl}}\right)^2\left(\frac{H_{end}}{m}\right)^{9/4} M_{pl}\sim 5\times 10^{-16}(1-n_s)^{13/4} M_{pl}.
\end{eqnarray}

\begin{remark}
Note that the formula of the reheating temperature is a little bit different from the one obtained in \cite{inflation1}, since there another thermalization rate has been used.
\end{remark}

By combining it with equation \eqref{eq47}, namely 
\begin{eqnarray}
Y-\frac{11}{4}\ln Y\cong 200.0,
\end{eqnarray}
one obtains that $n_s\cong 0.9721$, $r\cong 0.1117$ and $T_R\sim 11$ MeV, which means that this potential supports the production of heavy massive particles. 
If we consider massless particles far from the conformal coupling, then the reheating temperature becomes
\begin{eqnarray}
T_R\sim 3\times 10^{-2}\frac{H_E^2}{M_{pl}^2}M_{pl}\sim 3\times 10^{-10}(1-n_s)^2 M_{pl}\sim 5\times 10^5 \text{ GeV}.
\end{eqnarray}

Therefore, it falls out of the bounds that we have previously found ($1\text{ MeV}\leq T_R\leq 10^4$ GeV) and, thus, our model does not support the production of massless particles far from the conformal coupling.

\

Regarding the dynamics of the equation $\ddot{\varphi}+3H(\varphi,\dot{\varphi})\dot{\varphi}+V_{\varphi}=0$, we have also proceeded analogously as with the other potential. In Figure \ref{fig:dynamicspot3} we have represented some blue orbits solution of the dynamical system which start with infinite energy at $t\to-\infty$ and then evolve to the de Sitter phase for $t\to\infty$ after having gone through a phase transition at $\varphi=0$. In this case, the slow roll solution (in black) is $\dot{\varphi}=-\frac{V_{\varphi}}{3H}=M_{pl}^2\sqrt{\frac{4\lambda}{3}}$. We note that in this case the slow roll solution does not start with a de Sitter phase.

\

\begin{figure}[H]
\begin{center}
\includegraphics[scale=0.42]{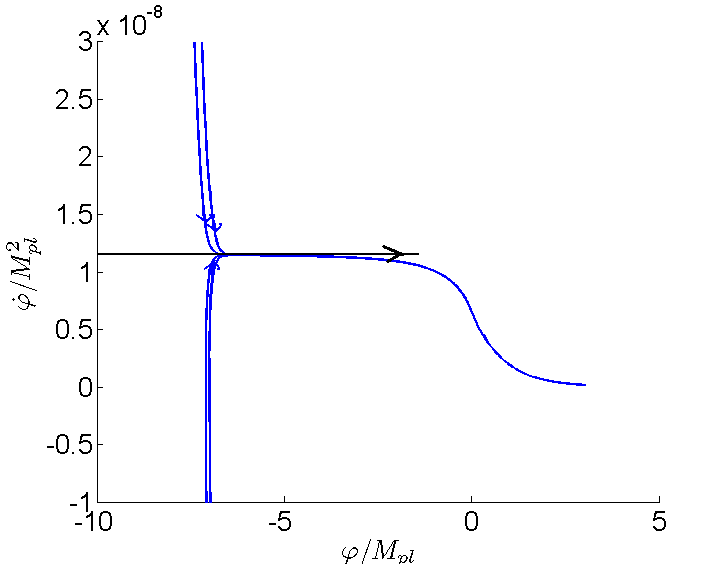}
\includegraphics[scale=0.42]{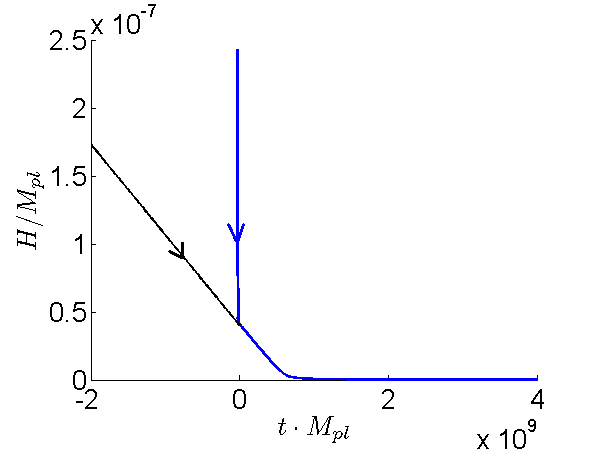}
\end{center}
\caption{Phase portrait in the plane $(\varphi/M_{pl},\dot{\varphi}/M_{pl}^2)$ (left) for some orbits, with the slow-roll one represented in black. Evolution of $H/M_{pl}$ in fuction of the time $t\times M_{pl}$ (right) for the same orbits represented in the phase portrait.}
\label{fig:dynamicspot3}
\end{figure}

\

We can compare this power law potential with the one studied in \cite{has}, namely
\begin{eqnarray}
V(\varphi)=\left\{ \begin{array}{ll} \lambda M_{pl}^2 (\varphi^2-\varphi_E^2), & \mbox{$\varphi<\varphi_E\not= 0$} \\ 0, & \mbox{$\varphi\geq \varphi_E\not= 0$.}  \end{array} \right. \label{potinflation3}
\end{eqnarray}

Firstly we note that for $\varphi\ll \varphi_E$, i.e during all the inflation, the behaviour will be analogous. Regarding the reheating temperature, the upper bound $T_R\leq 10^4$ GeV coincides. It only differs the fact that for heavy massive particles nearly conformally coupled with gravity 
the reheating temperature is in the GeV regime, instead of the MeV one, for the model coming from the potential in Equation \eqref{potinflation3}.

\section{Other quintessential inflation potentials}

In this last section we are going to study some other potentials that appear in \cite{enciclopedia} studying whether they can be implemented in the quintessential inflation.

\subsection{Open String Tachionic Inflation (OSTI) }

We consider the following adapted form of the OSTI potential \cite{KL}
\begin{eqnarray}
V(\varphi)=\left\{ \begin{array}{ll} -\lambda M_{pl}^2 \varphi^2\ln\left[\left(\frac{\varphi}{\varphi_0}\right)^2\right], & \mbox{$\varphi< 0 $} \\ 0, & \mbox{$\varphi\geq 0$}  \end{array} \right.
\end{eqnarray}
where $|\varphi_0|\gg M_{pl}$. We note that this potential is only non-negative for $\varphi>-|\varphi_0|$. Therefore, we will always move through this domain. Now, let's compute the cosmological parameters.
\begin{eqnarray}
\epsilon_*=2\left(\frac{M_{pl}}{\varphi_*}\right)^2\left(\frac{1+\ln\left[\left(\frac{\varphi_*}{\varphi_0}\right)^2\right]}{\ln\left[\left(\frac{\varphi_*}{\varphi_0}\right)^2\right]}\right)^2 \ \ \eta_*=2\left(\frac{M_{pl}}{\varphi_*}\right)^2\frac{3+\ln\left[\left(\frac{\varphi_*}{\varphi_0}\right)^2\right]}{\ln\left[\left(\frac{\varphi_*}{\varphi_0}\right)^2\right]}
\end{eqnarray}

Since $\epsilon$ cancels at $\varphi=-e^{-1/2}|\varphi_0|$, the inflation period will happen between this value and $\varphi_{end}$ (corresponding to $\epsilon=1$). Using that $|\varphi_0|\gg M_{pl}$, we obtain that $\varphi_{end}\approx \sqrt{2} M_{pl}$, $\varphi_*\approx\sqrt{\frac{8}{1-n_s}}M_{pl}$ and, hence, $N\approx \frac{2}{1-n_s}-\frac{1}{2}$, $r\approx 4(1-n_s)$ and $\alpha_s=-\frac{(1-n_s)^2}{2}$. Thus, a number of e-folds comprised between $63$ and $73$ corresponds to a spectral index $0.9685<n_s<0.9728$ and a ratio of tensor to scalar perturbations $0.109<r<0.126$. Therefore, the constraint $r\leq 0.12$ restricts this range  to $66\lesssim N\lesssim 73$, corresponding to $0.97\lesssim n_s\lesssim 0.973$, $0.11<r<0.12$ and $-4.5\times 10^{-4}\lesssim \alpha_s\lesssim -3.6\times 10^{-4}$. As usual, the power spectrum $P\approx -\frac{4\lambda}{3\pi^2(1-n_s)^2}\ln\left[\left(\frac{M_{pl}}{\varphi_0}\right)^2\frac{8}{1-n_s} \right]$ is imposed to be $P\approx 2\times 10^{-9}$ by choosing the suitable value of $\lambda$.

\

Now, we approximate the Hubble constant at the transition point by $H_E\sim H_{end}=\sqrt{\frac{V(\varphi_{end})}{2M_{pl}^2}}\approx 10^{-4}\sqrt{1-n_s}M_{pl}$. The combination of equation \eqref{eq13} and the number of e-folds for this potential leads to
\begin{eqnarray}
Y+\ln Y=216.2-\ln\left(\frac{g_R^{1/4}T_R}{\text{GeV}}\right) \label{eq54}
\end{eqnarray}
obtaining, as shown in Figure \ref{fig:rvsTosti}, that the valid bounds of the reheating temperature for this potential are $1 \text{ MeV}\leq T_R\leq 10^5$ GeV.

\begin{figure}[H]
\begin{center}
\includegraphics[height=45mm]{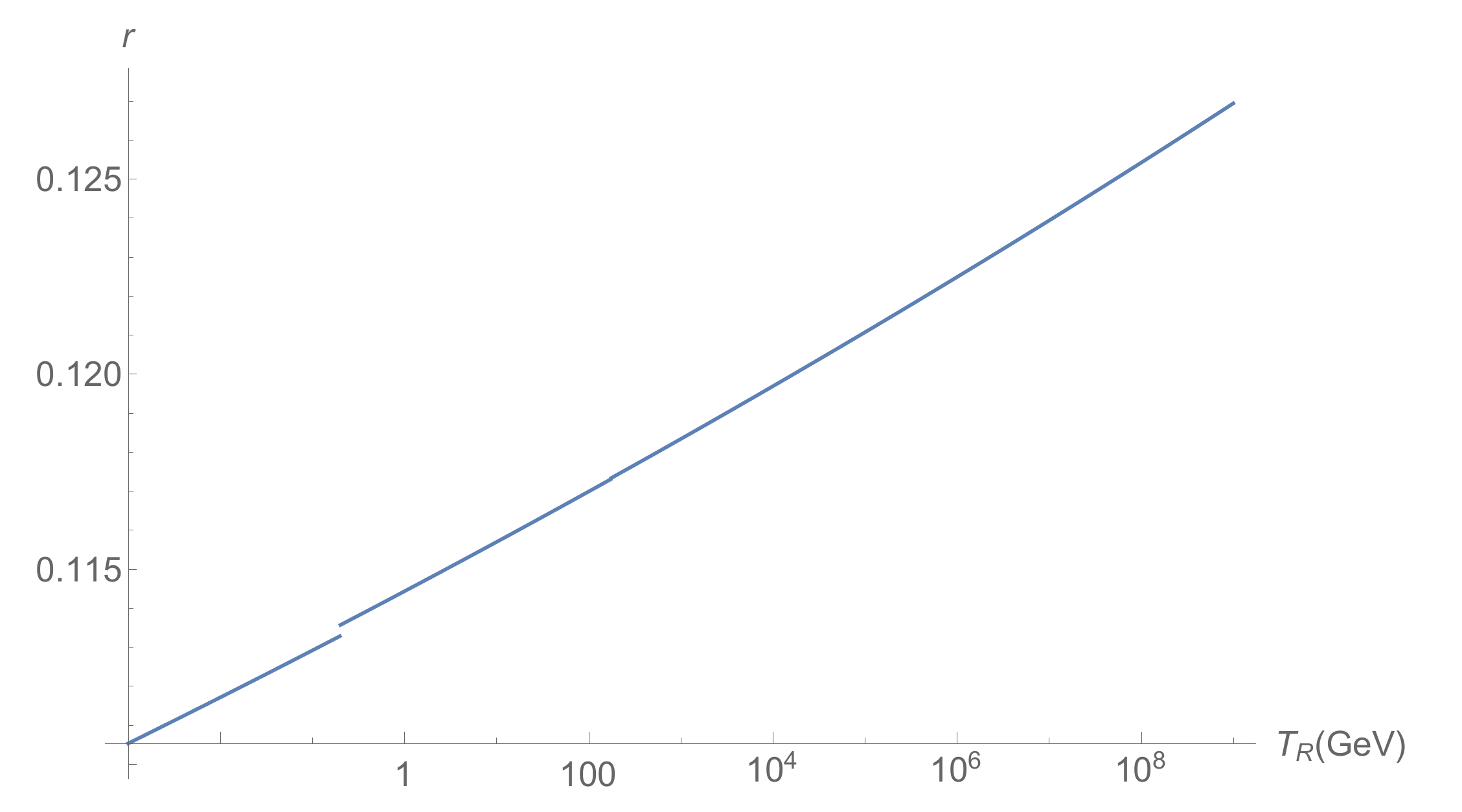}
\includegraphics[height=45mm]{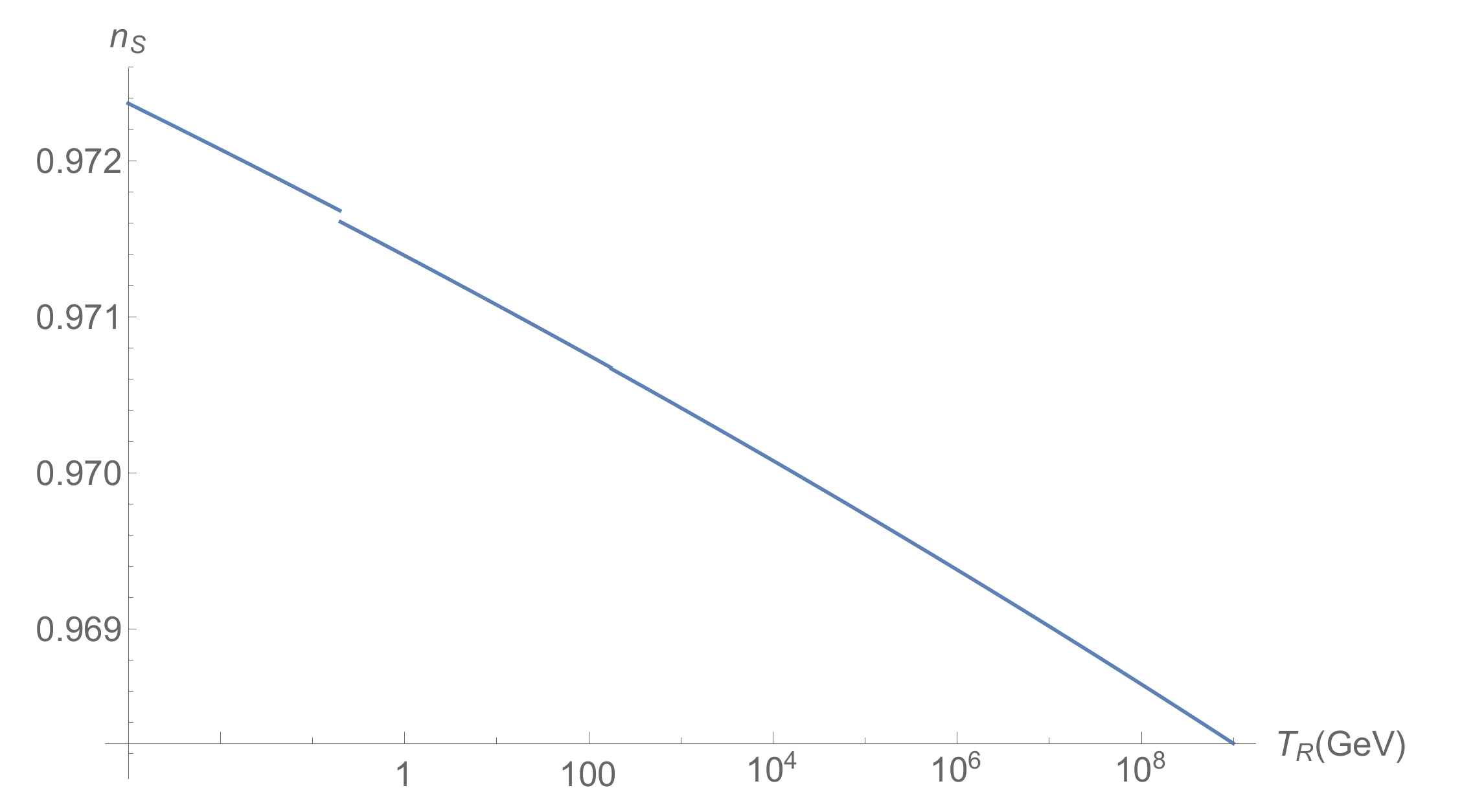}
\end{center}
\caption{Evolution of the tensor/scalar ratio $r$ (left) and the spectral index $n_s$ (right) versus the reheating temperature $T_R$.}
\label{fig:rvsTosti}
\end{figure}

As in the case of the power-law potential with $n=1$, this potential has a continuous derivative at the transition phase. Thus, when considering the production of massless particles nearly conformally coupled to gravity, the reheating temperature is
\begin{eqnarray}
T_R\sim \left|\xi-\frac{1}{6}\right|^{3/2}{\mathcal N}^{3/4}\frac{H_E^2}{M_{pl}}\approx 10^{-8}\left|\xi-\frac{1}{6}\right|^{3/2}(1-n_s)M_{pl}.
\end{eqnarray}

Thus, for $1 \text{ MeV}\leq T_R\leq 10^5$ GeV, we obtain that $10^{-8}\lesssim\left|\xi-\frac{1}{6}\right|\lesssim 3\times 10^{-3}$.  However, dealing with heavy massive particles ($m\sim M_{pl}$), one can see that 
the second derivative of the potential, and thus the third derivative of the Hubble parameter, diverges at the transition phase, which means that we cannot use the WKB solution to approximate the
modes. Therefore, we are not able to compute, in this case, the reheating temperature.
Regarding the case of massless particles far from the conformal coupling, the reheating temperature becomes
\begin{eqnarray}
T_R\sim 3\times 10^{-2}\frac{H_E^2}{M_{pl}^2}M_{pl}\sim 3\times 10^{-10}(1-n_s)M_{pl}\sim 2\times 10^7 \text{ GeV},
\end{eqnarray}
which shows that this model neither supports the production of these particles.

\

Finally, as done with the other potentials, we study the dynamics of the equation $\ddot{\varphi}+3H(\varphi,\dot{\varphi})\dot{\varphi}+V_{\varphi}=0$. In Figure \ref{fig:dynamicspotosti} we see again that the blue orbits (solution of the dynamics equation) start with infinite energy at $t\to-\infty$ and then evolve to the de Sitter phase after a phase transition at $\varphi=0$. As happened with the PLI potential, the slow roll inflation, namely $\dot{\varphi}= 2M_{pl}^2\sqrt{\frac{\lambda}{3}}\frac{\left|1+\ln\left[\left(\frac{\varphi}{\varphi_0} \right)^2 \right]\right|}{\sqrt{\left|\ln\left[\left(\frac{\varphi}{\varphi_0} \right)^2 \right]\right|}}$, does not start with a de Sitter phase.

\begin{figure}[H]
\begin{center}
\includegraphics[scale=0.35]{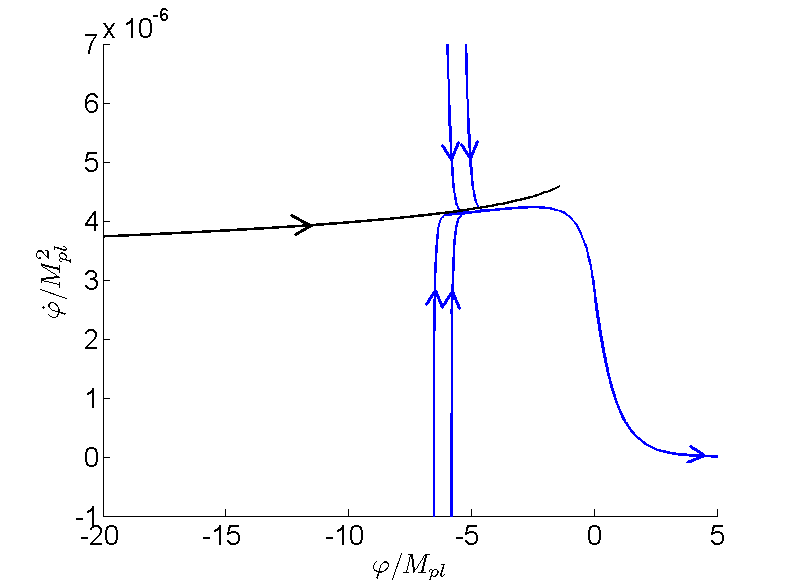}
\includegraphics[scale=0.41]{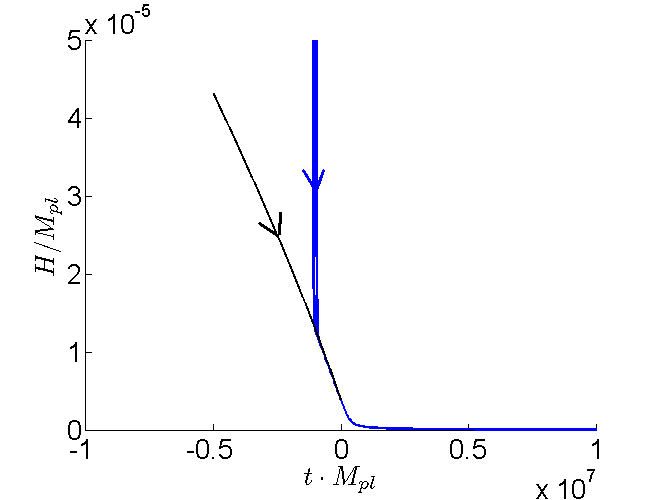}
\end{center}
\caption{Phase portrait in the plane $(\varphi/M_{pl},\dot{\varphi}/M_{pl}^2)$ (left) for some orbits, with the slow-roll one represented in black. Evolution of $H/M_{pl}$ in fuction of the time $t\times M_{pl}$ (right) for the same orbits represented in the phase portrait.}
\label{fig:dynamicspotosti}
\end{figure}

\subsection{Witten-O'Raifeartaigh Inflation (WRI)}
 
In this case, the version of the WRI \cite{Witten, R} potential is
\begin{eqnarray}
V(\varphi)=\left\{ \begin{array}{ll} \lambda M_{pl}^4\ln^2\left(\frac{-\varphi}{|\varphi_E|} \right), & \mbox{$\varphi< -|\varphi_E| $} \\ 0, & \mbox{$\varphi\geq -|\varphi_E|$}  \end{array} \right.,
\end{eqnarray}
where $|\varphi_E|\gg M_{pl}$. The cosmological parameters are
\begin{eqnarray}
\epsilon_*=2\left(\frac{M_{pl}}{\varphi_*\ln\left(\frac{-\varphi_*}{|\varphi_E|}\right)}\right)^2 \ \ \ \eta_*=2\left(\frac{M_{pl}}{\varphi_*}\right)^2\frac{\ln\left(\frac{-\varphi_*}{|\varphi_E|} \right)-1}{\ln^2\left(\frac{-\varphi_*}{|\varphi_E|} \right)}.
\end{eqnarray}

From this equation we have $\varphi_*=-|\varphi_E| e^{\frac{M_{pl}}{\varphi_*}\sqrt{\frac{2}{\epsilon_*}}}$. Since $|\varphi_*|\gg |\varphi_E|\gg M_{pl}$ and $\epsilon_*\sim 10^{-2}$ we have $\left|\frac{M_{pl}}{\varphi_*}\sqrt{\frac{2}{\epsilon_*}}\right|\ll 1 $, and thus,
$\varphi_*\cong -|\varphi_E| \left( 1+\frac{M_{pl}}{\varphi_*}\sqrt{\frac{2}{\epsilon_*}}\right)\cong -|\varphi_E| $.
Therefore, the approximation $\epsilon_*\approx 2\left(\frac{M_{pl}}{|\varphi|_E\ln\left(\frac{-\varphi_*}{|\varphi_E|} \right)}\right)^2\approx -\eta_*$ is valid and idem for $\epsilon_{end}$. Hence, we get that
\begin{eqnarray}
\varphi_*\approx -|\varphi_E|e^{\frac{4M_{pl}}{\sqrt{1-n_s}|\varphi_E|}} \ \ \ \ \varphi_{end}\approx -|\varphi_E|e^{\frac{\sqrt{2}M_{pl}}{|\varphi_E|}}
\end{eqnarray}

With regards to the number of e-folds, it can be exactly integrated as $N=\frac{|\varphi_E|^2}{8M_{pl}^2}\left.\left(x^2(2\ln(x)-1)\right)\right|_{x_*}^{x_{end}}$ (where $x=\frac{-\varphi}{|\varphi_E|}$), whose expression up to order 2 in $\frac{M_{pl}}{|\varphi_E|}$ is $N\approx \frac{4}{1-n_s}-\frac{1}{2}$, which results in a too high number of e-folds, namely $N\gtrsim 100$.


}

\

\subsection{K\"ahler Moduli Inflation I (KMII)}

The expression of the K\"ahler Moduli Inflation I (KMII) potential \cite{enciclopedia} is
\begin{eqnarray}
V(\varphi)=\left\{ \begin{array}{ll} \lambda M_{pl}^4\left(1-\alpha\frac{\varphi}{M_{pl}}e^{-\varphi/M_{pl}} \right), & \mbox{$\varphi>\varphi_E $} \\ 0, & \mbox{$\varphi\leq \varphi_E$}  \end{array} \right.,
\end{eqnarray}
where $\alpha$ is a positive dimensionless constant such that $\alpha\geq e$ and $\varphi_E$ is the value of $\varphi$ where $\frac{\varphi}{M_{pl}}e^{-\varphi/M_{pl}}=1/\alpha$ such that 
$\varphi_E\geq M_{pl}$. We note that, in contrast to the former potentials considered, in this one we are going to assume that $\dot{\varphi}<0$. The cosmological parameters are
\begin{eqnarray}
\epsilon=\frac{1}{2}\frac{\alpha^2(x-1)^2 e^{-2x}}{(1-\alpha xe^{-x})^2} \ \ \ \eta=\frac{\alpha(2-x)e^{-x}}{1-\alpha xe^{x}}
\end{eqnarray}
where $x=\varphi/M_{pl}$. By considering that $x_*\gg 1$, we obtain that $n_s-1\approx2\alpha x_*e^{-x_*}$. Regarding the number of e-folds, they can be exactly integrated, namely
\begin{eqnarray}
N=x_{end}-x_*+\ln\left(\frac{x_{end}-1}{x_*-1}\right)+\frac{e}{\alpha}\left(Ei(x_*-1)-Ei(x_{end}-1) \right),
\end{eqnarray}
where $x_{end}$ is the value of $x$ where $\epsilon=1$ and $E_i$ is the exponential integral function, which verifies that for $x_*\gg 1$, $E_i(x_*-1)\approx \frac{e^{x_*}}{x_*}$, being this the dominant term in the previous equation. Hence, by using that $x_*e^{-x_*}\approx\frac{n_s-1}{2\alpha}$, one obtains that
\begin{eqnarray}
N\approx \frac{2e}{1-n_s},
\end{eqnarray}
which does not fulfill our bounds for the number of e-folds and the spectral index. Therefore, we have proved that this potential is not a viable quintessential inflation model.

\subsection{Brane Inflation (BI)}

The Brane Inflation (BI) potential behaves as \cite{lmr}
\begin{eqnarray}
V(\varphi)=\left\{ \begin{array}{ll} \lambda M_{pl}^4\left[1-\left(\frac{-\varphi}{\mu M_{pl}}\right)^{-p} \right], & \mbox{$\varphi<\varphi_E$} \\ 0, & \mbox{$\varphi\geq \varphi_E$}  \end{array} \right.,
\end{eqnarray}
where $\mu$ and $p$ are positive dimensionless parameters and $\varphi_E\equiv -\mu M_{pl}$ 
We will consider for simplicity only $p\in\mathbb{N}$. Thus, the cosmological parameters are
\begin{eqnarray}
\epsilon=\frac{1}{\mu^2}\frac{p^2}{2x^2(1-x^p)^2} \ \ \eta=\frac{p(p+1)}{\mu^2x^2(1-x^p)},
\end{eqnarray}
where $x=\frac{-\varphi}{\mu M_{pl}}$. We are going to differ two cases:

\begin{itemize}

\item[a)] $\mu\ll 1$:

In this case one will have that $x_*\gg 1$, and thus $\epsilon_*\cong \frac{p^2}{2\mu^2 x_*^{2(p+1)}}$ and 
$\eta_*\cong \frac{p(p+1)}{\mu^2 x_*^{p+2}}$, meaning that $n_s-1\cong 2\eta_*\cong -\frac{2p(p+1)}{\mu^2 x_*^{p+2}}$. On the other hand,  one has
$$N=\frac{\mu^2}{p}\left[\frac{x_*^{p+2}}{p+2}-\frac{x_{end}^{p+2}}{p+2}-
\frac{x_*^2}{2}+\frac{x_{end}^2}{2}\right].$$

Taking into account that $x_{end}\cong \left( \frac{p^2}{2\mu^2} \right)^{\frac{1}{2(p+1)}}\ll x_*$, one has $N \cong \frac{\mu^2}{p}\frac{x_*^{p+2}}{p+2}$, meaning that
$N\approx \frac{2(p+1)}{(1-n_s)(p+2)}$ which enters in our range for values of  $p$ greater than $17$.
For the tensor/scalar ratio one has $r\cong \frac{8p^2}{\mu^2 x_*^{2(p+1)}}\ll \frac{8p^2}{\mu^2 x_*^{p+2}}$ for $p\geq 1$, namely $r\ll \frac{4 p(1-n_s)}{p+1}\leq 4(1-n_s)$. Hence, one can conclude that for all the values of $2\sigma$ CL of the spectral index it is verified that $r<0.12$. We find as well that $\alpha_s\approx -\frac{(1-n_s)^2}{2}$, which also fulfills the constraints for the running. And finally, by adjusting $\lambda$ so that $P\sim 2\times 10^{-9}$, we can build a successful quintessential model. 

\

Effectively, regarding the reheating constraints we obtain that for all the restricted values of the parameter $p$, $H_E\sim 10^{-4}\sqrt{1-n_s}M_{pl}$. So, as usual, one obtains that the reheating temperature bounds from nucleosynthesis give the constraint $0.968\leq n_s\leq 0.972$. For massive particles we have that $T_R\sim 10^3$ GeV. In the case of massless particles nearly conformally coupled with gravity, we obtain that $\left|\xi-\frac{1}{6}\right|\gtrsim 10^{-8}$ and the fact that $\left|\xi-\frac{1}{6}\right|\lesssim 1$ should be satisfied constraints our reheating temperature to be less than $10^7$ GeV. Finally, considering massless particles far from conformal coupling, one finds that $T_R\sim 10^7$ GeV.

\

\item[b)] $\mu\gg 1$:

In this case by taking for example $\mu\geq 100p$, since $\epsilon_*$ or $\eta_*$ have to be of the order of $10^{-2}$, it is verified that $x_*\cong 1$ and, thus,
\begin{eqnarray}
\epsilon_*\cong\frac{1}{\mu^2}\frac{p^2}{2(1-x_*^p)^2} \ \ \eta_*\cong\frac{p(p+1)}{\mu^2(1-x_*^p)},
\end{eqnarray}
So, given that $\epsilon_*\gg \eta_*$, $1-n_s\cong 6\epsilon_*\cong \frac{1}{\mu^2}\frac{3p^2}{(1-x_*^p)^2}$. { This means that $x_*\cong 1-\sqrt{\frac{3}{1-n_s}}\frac{1}{\mu}$, as well as 
$x_{end}\cong 1-\sqrt{\frac{1}{2}}\frac{1}{\mu}$. Consequently,
$N\sim \frac{1}{2}\left(\frac{3}{1-n_s} -1 \right)\sim 50$, which does not enter in our range.}











\end{itemize}


\subsection{Loop Inflation (LI)}

In this case the potential behaves as \cite{Dvali}
\begin{eqnarray}
V(\varphi)=\left\{ \begin{array}{cc}
0 ,& \varphi\leq \varphi_E\equiv M_{pl}e^{-\frac{1}{\alpha}}\\
\lambda M_{pl}^4\left(1+\alpha \ln \left(\frac{\varphi}{M_{pl}}\right) \right),& \varphi\geq \varphi_E\equiv M_{pl}e^{-\frac{1}{\alpha}},
\end{array}\right.
\end{eqnarray}
where $\lambda$ and $\alpha$ are positive dimensionless constants. The slow roll parameters are
\begin{eqnarray}
\epsilon=\frac{1}{2x^2}\frac{\alpha^2}{\left(1+\alpha \ln x  \right)^2}, \qquad
\eta=-\frac{1}{x^2}\frac{\alpha}{\left(1+\alpha \ln x \right)},
\end{eqnarray}
where we have introduced the parameter $x\equiv \frac{\varphi}{M_{pl}}$.

We consider to different asymptotic cases:
\begin{enumerate}
\item $0<\alpha \ll 1$:

In this case one has 
\begin{eqnarray}
\epsilon_*\cong\frac{\alpha^2}{2x^2_*}, \qquad
\eta_*\cong-\frac{M_{pl}^2\alpha}{\varphi^2_*},
\end{eqnarray}
and thus, $n_s-1\cong -2\eta_*\cong \frac{2\alpha}{x^2_*}$. For the number of e-folds one has
\begin{eqnarray}
N\cong \frac{x_*^2}{2\alpha}\cong \frac{1}{1-n_s},
\end{eqnarray}
which leads, as in the case of HI, to a not high enough number of e-folds.

 \item $\alpha \gg1$:
 
 Now the slow roll parameters become
 \begin{eqnarray}
\epsilon_*=\frac{1}{2x^2_*\ln^2 x_*}, \qquad
\eta_*=-\frac{1}{x^2_*\ln x_*},
\end{eqnarray}
and the spectral index and the tensor/scalar ratio will be as a function of $x_*$
\begin{eqnarray}
1-n_s=\frac{1}{x^2_*\ln^2 x_*}(3+2\ln x_*),\qquad r=\frac{8}{x^2_*\ln^2 x_*}.
\end{eqnarray}

Then, at $2\sigma$ C.L., for the allowed values of the spectral index, we can see, after some numerics, that $x_*$ ranges in the domain $6.94\leq x_*\leq 7.98$. On the other hand, the number of e-folds is 
\begin{eqnarray}\label{eq76}
N=\frac{x_*^2}{2}\left(\ln x_* -\frac{1}{2}\right)-\frac{x_{end}^2}{2}\left(\ln x_{end}-\frac{1}{2}\right).
\end{eqnarray}

Using the range of values for $x_*$ one finds that $34\lesssim N\lesssim 50$, which comes out of the nucleosynthesis bounds.

\end{enumerate}
 
 \section{Conclusions}
We have  adapted some inflationary potentials to quintessence inflation, extending them to zero after they vanish and adding a small cosmological constant. Once we have done it, we have tested the models imposing that: 1.- they fit well with the current observational data provided by BICEP and Planck teams. 2.- The number of e-folds  must range between $63$ and $73$, this number is larger than the usual one used for potentials with a deep well, due to the kination phase after inflation. 3.- The reheating temperature due to the gravitation particle production during the phase transition from inflation to kination has to be compatible with the nucleosynthesis success, i.e., it has to range between $1$ MeV and $10^9$ GeV, although if one wants to remove the gravitino and moduli problems, { which appear in
supergravity}, this temperature has to be less than $1$ GeV.

\

Our study shows that the potentials WRI and KMII lead to a too high number of e-folds, while for HI and LI potentials this number is too small. Other potentials such as ESI, PLI (only when the potential is quadratic), OSTI and BI satisfy the prescriptions $1$ and $2$. 
Dealing with the reheating temperature, all the four  models, namely ESI, PLI, OSTI and BI, lead to a temperature compatible with the nucleosynthesis bounds. However, since the ESI potential comes from a supersymmetric gravitational one, the problems related with the gravitino and moduli overproduction will appear, which can only be removed when the reheating temperature is in the MeV regime. Then, for this potential, when reheating is due to the production of massless nearly conformally coupled particles (the coupling constant is very close to $1/6$), these problems are removed because the reheating temperature is very low (in the MeV regime). On the contrary, reheating due to the production of very heavy massive particles conformally coupled with gravity or massless particles far from the conformal coupling leads to a too high reheating temperature, which does not avoid the gravitino and moduli problems for the ESI potential.

\

Fortunately, the other three potentials, namely PLI, OSTI and BI, do not come from supersymmetric potentials and, thus, do not suffer these problems because gravitino and moduli fields only appear in supergravity, meaning that the reheating via very massive particles conformally coupled with gravity or via massless particles far from the conformal coupling is also viable for these three potentials.

\section*{Acknowledgments} 
We would like to thank Professor Sergei D. Odintsov for reading the manuscript. This investigation has
been supported in part by MINECO (Spain), Project 
No. MTM2014-52402-C3-1-P.

{
\section*{Appendix: A critical review of the gravitational wave particle production} 

Gravitons satisfy the same equation as scalar fields minimally coupled with gravity,
so its energy density 
at the end of inflation is twice that of a single scalar
field due to  the two graviton polarization states. Experimental observations show that, at the end of inflation, the ratio of the energy density of the relic gravitational waves to the one of the produced particles is less than $10^{-2}$ \cite{{ghmss}}. In \cite{pv},  using the results of \cite{dv} and \cite{Giovannini},  the authors take as the energy density of the gravitons 
$\rho_g=2\times 10^{-2} H_E^4$, where $H_E$ is the value of the Hubble parameter at the transition time, so assuming that there are ${\mathcal N}_s$ different light  scalar fields with a coupling constant near zero, one deduces that $\frac{\rho_g}{\rho_{\chi}}=\frac{2}{{\mathcal N}_s}$, where $\rho_{\chi}$ denotes the energy density of the produced particles. Then, according to the observational bound, this means that the number of different scalar fields must be greater than $100$, which only happens in some supersymmetric theories \cite{pv}.

\

Note that in this situation the reheating temperature is of the order $10^3 {\mathcal N}_s^{3/4}$ GeV $\sim 3\times 10^4$ GeV, allowing the overproduction of gravitinos and moduli fields. 
The problem is even worse when one considers reheating via the production of heavy massive particles because this mechanism is less efficient than the one due to the creation of massless particles.
The situation could be addressed when one considers models whose potential does not come from supergravity, which avoids the gravitino and moduli problems, and reheating is produced via oscillations of an
auxiliary field in hibrid quintessential models \cite{Berera} or via instant preheating \cite{sss, ghmss, hmss} because these kinds of reheating are more efficient  than the one due to gravitational particle production.

\

However, a more detailed investigation shows that the results used in \cite{pv} about the production of gravitons may not be correct and the conclusions could be wrong. Here, we summarize the problems that appear in this calculation

\

 1.- Gravitational particle production of massless particles far from the conformal coupling is only considered in toy models, where there is a transition from the exact de Sitter phase to
a kination or radiation one. This happens because the vacuum modes are represented by Hankel functions and one can perform analytic calculation, but, as it is well known, the inflationary period is not exactly de Sitter. 

\

 2.- In the calculations, only the energy density of the particles is considered and vacuum polarization effects  (see \cite{glavan} for a detailed calculation) are disregarded, but as has been showed in \cite{fkl}  they could be very important in some inflationary scenarios leading to undesirable effects.

 \
 
 3.- Performing the calculations, an UV divergence appears but it is removed by disregarding the modes inside the Hubble horizon, because they ``do not feel" gravity and, thus, do not produce particles, but in the case of gravitons we obtain an IR divergence which is ignored and produces important consequences. Let's explain it in detail:

 \
 
When one considers a transition from the de Sitter phase to radiation, in \cite{dv} it was showed that the density energy of the produced massless particles coupled with gravity, in the
 long wave-length  limit, i.e., for modes that leave the Hubble radius after the phase transition ($k\leq a_E H_E$, being $H_E$ the Hubble parameter during the de Sitter phase and $a_E$ the scale factor at the transition time), satisfies
\begin{eqnarray}
d\rho_{\chi}=\frac{\Gamma^2(\nu)}{32\pi^3}\left(\frac{a_E}{a(t)} \right)^4\left( \nu-\frac{1}{2} \right)^2\left(\frac{2H_E}{k}  \right)^{2\nu+1}k^3dk,
\end{eqnarray}
 where $\nu^2=\frac{9}{4}-12\xi$, being $\xi$ the coupling constant. This quantity is IR divergent for $\nu\geq 3/2$, that is, for $\xi\leq 0$, and in this case, to obtain the energy density of the produced particles, one cannot simply integrate in the domain 
 $[0, a_E H_E]$. One can only do that when $\xi >0$, obtaining the well-known result $\rho\sim 10^{-2} H_E^4 \left(\frac{a_E}{a(t)} \right)^4$.  Thus,  the case of gravitons has to be considered in another way, because for them $\xi=0$. This was done in \cite{haro}, where the IR singularity is removed assuming an early radiation phase before inflation (see \cite{fp}). Then, let $a_i$ be the scale factor when the universe passes from the early radiation phase to inflation. In this situation, if one is only interested in the gravitons produced by the modes that leave the Hubble radius during
 the de Sitter phase, i.e., in modes satisfying $ a_i H_E\leq k\leq a_E H_E$, one obtains \cite{haro}
\begin{eqnarray}
\rho_{g}=\frac{H_E^4}{4\pi^2}\left(\frac{a_E}{a(t)} \right)^4\ln\left( \frac{a_E}{a_i} \right)=\frac{H_E^4{\mathcal N}}{4\pi^2}\left(\frac{a_E}{a(t)} \right)^4,\end{eqnarray}
where ${\mathcal N}\equiv \ln\left( \frac{a_E}{a_i} \right)$ is the number of e-folds that the de Sitter phase lasts. On the other hand, dealing with massless particles with a small negative coupling constant, one has
\begin{eqnarray}
\rho_{\chi}=\frac{4^{\nu}}{16\pi^3(2\nu-3)}\Gamma^2(\nu)\left( \nu-\frac{1}{2} \right)^2 H_E^{2\nu+1} \left(\frac{a_E}{a(t)} \right)^4
\left( e^{(2\nu-3){\mathcal N}}-1  \right)\cong \frac{ e^{(2\nu-3){\mathcal N}}-1}{8\pi^2(2\nu-3)} H_E^4 \left(\frac{a_E}{a(t)} \right)^4,
\end{eqnarray}
and thus,
\begin{eqnarray}
\frac{\rho_g}{\rho_{\chi}}\cong \frac{2(2\nu-3){\mathcal N}}{e^{(2\nu-3){\mathcal N}}-1}\cong \frac{16|\xi|{\mathcal N}}{e^{8|\xi|{\mathcal N}}-1},\end{eqnarray}
where we have used that for negative small values of the coupling constant it is satisfied that $\nu=\frac{3}{2}+4|\xi|$. Since ${\mathcal N}$ has to be greater than $N$ (the number of e-folds from the leaving of the pivot scale to the end of inflation), one can safely take ${\mathcal N}\geq 100$, then choosing for example ${\mathcal N}=100$, and $\xi=-0.012$ the bound is reached. So, a unique massless quantum field with a small negative coupling constant satisfies the bound $\frac{\rho_g}{\rho_{\chi}}\leq 10^{-2} $, meaning that gravitational supersymmetric potentials are not needed and 
the gravitino and moduli problems do not appear, allowing a greater reheating temperature. In this case, the reheating temperature, taking as usual $H_E\sim 10^{-6} M_{pl}$, is given by $T_R\sim
2\times 10^4 \left(\frac{e^{8|\xi|{\mathcal N}}-1}{|\xi|}\right)^{3/4}$ GeV, leading for the values ${\mathcal N}=100$ and $\xi=-0.012$ to the reheating temperature $T_R\sim 7\times 10^8$ GeV, which enters in the 
viable range. 

\

Summing up, the problem of the production of massless particles far from the conformal coupling and in particular the overproduction of gravitational waves in quintessential inflation is a problem that deserves future investigation for many reasons, but basically because nowadays we do not have any analytic method which allows us to calculate, in realistic models,  
the backreaction of a massless quantum field not conformally coupled with gravity,
and only results that are valid for toy models which may not apply to viable models are used to perform analytic calculations.
}

\end{document}